\documentclass[twocolumn]{emulateapj}
\pdfoutput=1
\usepackage[]{units}
\usepackage{datetime}
\usepackage{graphicx}
\usepackage{amssymb}
\usepackage{amsmath}
\usepackage{threeparttable}
\usepackage[bookmarks=false]{hyperref}
\usepackage[usenames,dvipsnames]{xcolor}

\newcommand{\hst}{{\it HST}}

\newcommand{\dg}{\ensuremath{^{\rm o}}}
\mathchardef\mhyphen="2D
\slugcomment{Submitted to MNRAS Sept 19, 2015}
\shorttitle{Giant scattering cones}
\shortauthors{Obied et al.}

\begin{document}

\title{Giant scattering cones in obscured quasars}

\author{Georges Obied\altaffilmark{1,2}, Nadia L. Zakamska\altaffilmark{1}, Dominika Wylezalek\altaffilmark{1}, Guilin Liu\altaffilmark{3}}
\altaffiltext{1}{Department of Physics \& Astronomy, Johns Hopkins University, Bloomberg Center, 3400 N. Charles St., Baltimore, MD 21218, USA}
\altaffiltext{2}{Department of Physics, Harvard University, 17 Oxford Street, Cambridge MA 02138, USA}
\altaffiltext{3}{Department of Physics, Virginia Tech, Blacksburg, VA 24061, USA}

\begin{abstract}
We analyze {\it Hubble Space Telescope} observations of scattering regions in 20 luminous obscured quasars at $0.24<z<0.65$ (11 new observations and 9 archival ones) observed at rest-frame $\sim 3000$\AA. We find spectacular $5-10$ kpc-scale scattering regions in almost all cases. The median scattering efficiency at this wavelength (the ratio of observed to estimated intrinsic flux) is 2.3\%, and 73\% of the observed flux at this wavelength is due to scattered light, which if unaccounted for may strongly bias estimates of quasar hosts' star formation rates. Modeling these regions as illuminated dusty cones, we estimate the radial density distributions of the interstellar medium as well as the geometric properties of circumnuclear quasar obscuration -- inclinations and covering factors. Small derived opening angles (median half-angle and standard deviation 27\dg$\pm$9\dg) are inconsistent with a 1:1 type 1 / type 2 ratio. We suggest that quasar obscuration is patchy and that the observer has a $\sim 40\%$ chance of seeing a type 1 source even through the obscuration. We estimate median density profile of the scattering medium to be $n_{\rm H}=0.04-0.5$ $(1{\rm kpc}/r)^2$ cm$^{-3}$, depending on the method. Quasars in our sample likely exhibit galaxy-wide winds, but if these consist of optically thick clouds then only a small fraction of the wind mass ($\la 10\%$) contributes to scattering.
\end{abstract}

\keywords{galaxies: ISM -- polarization -- quasars: general -- scattering}

\section{Introduction}
\label{sec:intro}

The unification model proposed by \citet{anto93} holds that active galactic nuclei (AGN) are intrinsically the same and that differences in their observed spectral properties are due to different relative orientations of the nucleus and surrounding opaque dust with respect to the observer. An active nucleus possesses a compact strong radiation source (presumably an accretion disk around the supermassive black hole and a surrounding corona), as well as an emission-line region with characteristic velocities of a few thousand km s$^{-1}$. If we have a direct view to the central engine, then we observe an unobscured, ``type 1'' source, with characteristic strong ultra-violet, optical and X-ray radiation from the accretion disk and broad emission lines. If the direct view to the central engine is blocked by circumnuclear dust, then a hidden ``type 2'' nucleus may be identified using indirect signatures, such as strong infrared radiation produced by the circumnuclear dust or narrow emission lines produced in photo-ionized regions outside obscuration which have a direct view to the nucleus.

The unification model was developed and tested for nearby relatively low-luminosity AGN -- Seyfert galaxies -- using imaging, polarimetry and spectropolarimetry of scattered-light regions \citep{anto85, bail88, mill90, mill91, tran92, tran95a, tran95b, tran95c, cape95, kish99}. Even if the direct view to the nucleus is obscured, the quasar light can escape along other unobscured directions, scatter off surrounding material and reach the observer. The scattered component can be then identified either via its polarization signature or its morphology in imaging observations, as illumination of extended material by a light source blocked along some directions produces a characteristic conical shape.

Whether the same orientation-based unification model is directly applicable to quasars -- high-luminosity ($L_{\rm bol}\ga 10^{45}$ erg s$^{-1}$) active nuclei -- is still unclear. Even the basic measurement of the ratio of obscured to unobscured quasars remains problematic \citep{lawr10}. For quite a while, few obscured quasars were known, leading to suggestions that the powerful radiation of a luminous quasar obliterates obscuring dust out to large distances and that the obscured fraction might as a result decline with luminosity (so-called `receding torus' model). These ideas found support in X-ray surveys \citep{ueda03, stef03, lafr05, barg05, hasi08} which uncovered few obscured quasars at high luminosities and, indirectly, in the spectral energy distribution of luminous quasars \citep{trei08}. At the same time, wide-field infrared, optical and radio surveys uncovered large numbers of obscured quasar candidates whose number densities are similar to those of unobscured quasars at the same luminosity \citep{zaka03, lacy04, ster05, mart06, glik07, reye08, donl12, asse15, lacy15}. Thus a full accounting of the obscured quasar population is still incomplete. 

Furthermore, the physical nature and the dynamical state of the obscuring material remains poorly understood \citep{krol07}. Despite many questions about the origins and the long-term stability of an `obscuring torus', this concept remains popular, in part because it finds strong support in observations. Indeed, classical polarimetric observations of low-redshift luminous obscured quasars confirm that some of these objects would be seen as unobscured type 1 sources if viewed along another direction, and a number of very extended ($\ga 10$ kpc) conical scattered light nebulae have been detected in such sources \citep{hine93, hine95, hine99, tran00, smit03, zaka05, zaka06, schm07, borg08}. 

On the other hand, obscured quasars -- in particular those at high redshifts -- increasingly present with a confusing mix of signatures that are no longer well explained by the standard unification model \citep{alex13, asse15, ross15, tsai15}. Perhaps some of these objects represent the young dust-enshrouded phase of quasar obscuration \citep{hopk06}, or perhaps extinction on the scales of the entire galaxy is important \citep{lacy07}, especially at high redshifts. Observations of scattered light in quasars of different luminosities and redshifts will help probe the different models of quasar obscuration.

Because a luminous quasar is capable of illuminating and ionizing interstellar medium out to large distances, scattering in luminous quasars occurs on scales of several kpc. Thus polarimetric and scattered light observations not only offer a direct test of the geometric unification model of quasars, but also provide a unique probe of the physical conditions in the interstellar medium of the quasar host galaxy. In this paper we present observations, modeling and physical parameters of giant scattering regions around luminous obscured quasars. In Section \ref{sec:data} we describe observations and data reduction. In Section \ref{sec:model}, we identify and model scattered light regions. In Section \ref{sec:discussion} we discuss the implications of our measurements and we summarize in Section \ref{sec:conclusions}. We use a $h$=0.7, $\Omega_m$=0.3, $\Omega_{\Lambda}$=0.7 cosmology throughout this paper. Although both the {\it Hubble Space Telescope} and the Sloan Digital Sky Survey use vacuum wavelengths, we use air wavelengths for designating emission lines following long-standing convention.

\section{Sample selection, observations and data reductions}
\label{sec:data}

Type 2 quasars studied in this paper are drawn from the optically-selected type 2 quasar candidates presented by \citet{zaka03} and \citet{reye08}. This large parent sample is selected from the first generation of the spectroscopic part of the Sloan Digital Sky Survey (SDSS; \citealt{york00}) based on emission-line properties. Briefly, type 2 quasar candidates at $z\la 1$ are required to show narrow (full width at half maximum $<2000$ km s$^{-1}$) emission lines with line ratios characteristic of photo-ionization by a hidden quasar continuum. Because the full range of standard emission line diagnostics \citep{bald81} is inaccessible in the optical spectrum for $z\ga 0.3$, we require a high [OIII]$\lambda$5007\AA/H$\beta$ ratio (at least 3, but in practice $\sim 10$ for the luminous type 2 quasars discussed here) accompanied by additional signatures to distinguish these objects from low-metallicity star forming galaxies (e.g., high-ionization emission lines such as [NeV]$\lambda\lambda$3346,3426\AA). Targeting of such objects for follow-up spectroscopy is incomplete and is often done on the basis of unusual colors or presence of a faint radio counterpart \citep{reye08}.

Extensive multi-wavelength follow-up studies of the samples of \citet{zaka03} and \citet{reye08} by our group and others support an overall picture in which luminous quasars are hidden from the observer by circumnuclear gas and dust, as high column densities of intervening material are directly detected in X-ray observations, with roughly half of the objects being Compton-thick \citep{ptak06, vign10, jia13, lans14, lans15}. While due to the obscuration these objects are relatively faint at ultra-violet and optical wavelengths, they show high (up to $10^{47}$ erg s$^{-1}$) infrared luminosities presumably due to the thermal re-radiation of the quasar emission by the obscuring dust \citep{zaka04, zaka08, mate13, zaka15}. The fraction of radio-loud objects \citep{zaka04, lal10, zaka14} is similar to that of unobscured quasars, suggesting that the incidence and detectability of powerful jets from supermassive black holes is roughly independent of the geometry of obscuration. Finally, spectropolarimetry of type 2 quasar candidates reveals polarized broad emission lines typical of unobscured (type 1) quasars, directly confirming that these objects would be seen as type 1s along some other lines of sight \citep{zaka05}.

Type 2 quasars are ideally suited for studies of hosts of luminous quasars as the circumnuclear obscuration provides a natural coronagraph and an opportunity to observe the host galaxy without the bright glare of the quasar itself. We therefore conducted follow-up {\it Hubble Space Telescope} imaging of some of type 2 quasar candidates on two occasions. In 2003 -- 2004, nine radio-quiet type 2 quasars were selected on the basis of their high [OIII] luminosity from the original parent sample of about 150 type 2 quasars of \citet{zaka03}; these \hst\ observations (GO-9905, PI Strauss) were first presented by \citet{zaka05,zaka06} and are re-analyzed here.

\begin{figure*}
\centering
\includegraphics[scale=0.7, clip=true, trim=0cm 10cm 10.5cm 0cm]{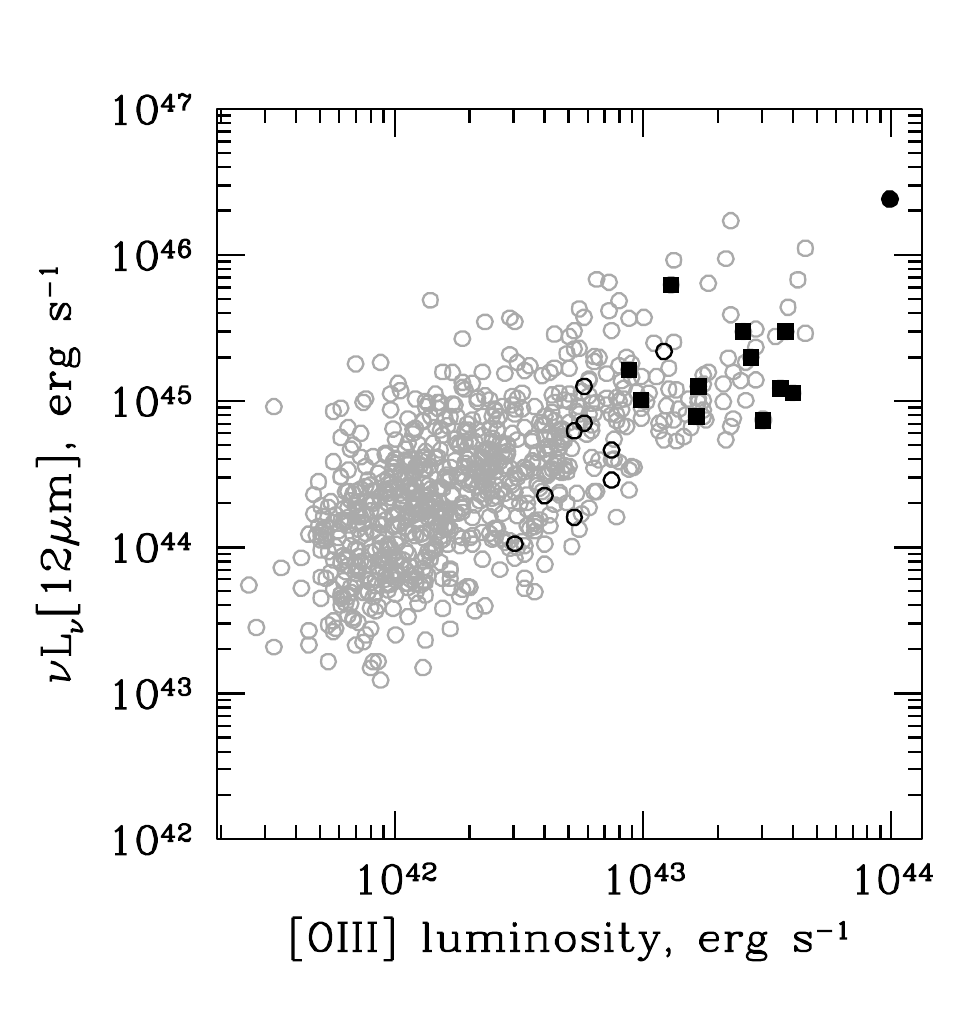} 
\includegraphics[scale=0.7, clip=true, trim=0cm 10cm 10.5cm 0cm]{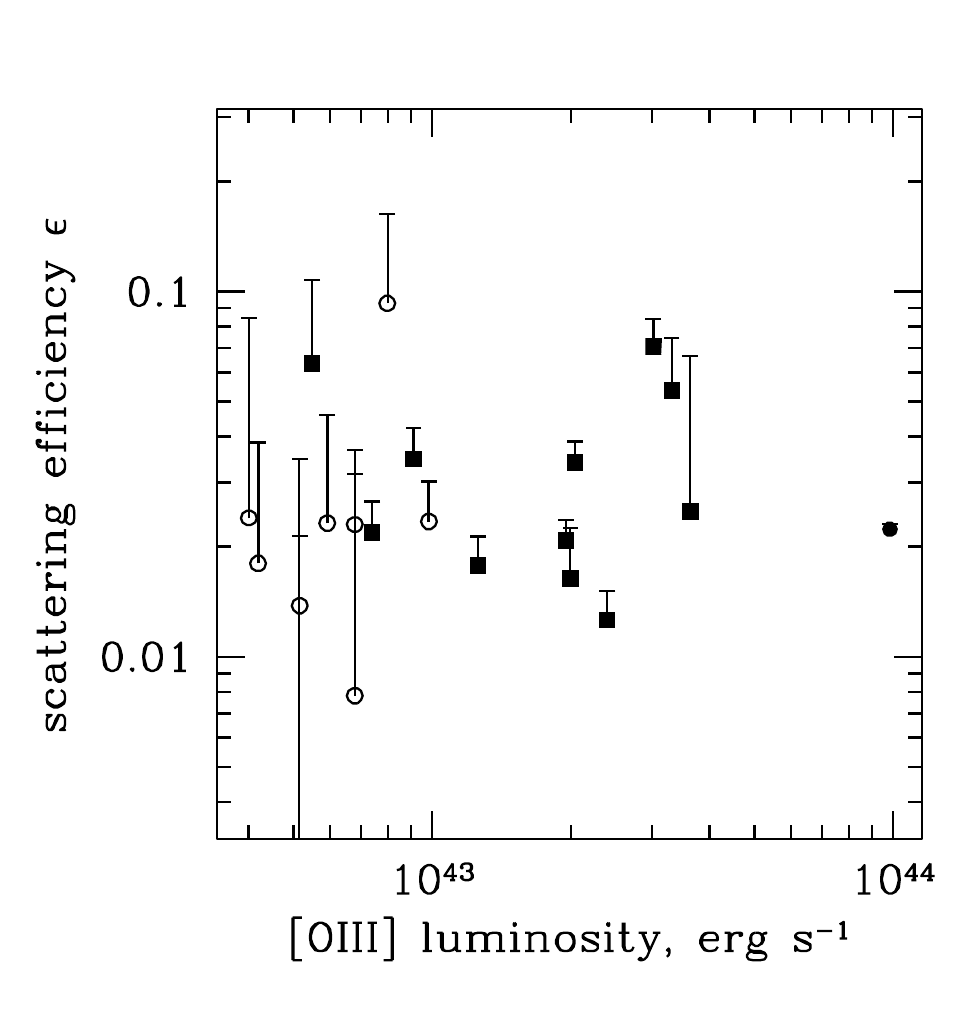}
\caption{Left: Line luminosities and 12\micron\ infrared luminosities (from {\it WISE}) for the archival \hst\ sample (open circles, from \citealt{zaka03, zaka05, zaka06}), new \hst\ sample (filled squares, from \citealt{liu13a, liu13b}) and IRAS 09104+4109 (filled circle, from \citealt{hine99}) compared to SDSS type 2 quasars (grey points; \citealt{reye08}). Right: scattering efficiencies at 3000\AA\ -- the ratio of observed to estimated intrinsic luminosity at this wavelength. The points show the scattering efficiencies calculated based on our best stellar subtraction, whereas the error bars show the upper limits on the scattering efficiencies if no stellar subtraction is performed and all of the observed $U$-band flux is attributed to scattered light. }
\label{pic_example}
\end{figure*}

As the SDSS progressed and the sample was expanded to $\sim 900$ sources \citep{reye08}, we were able to extend the luminosity range to higher values of $L$[OIII]. A further subsample of 11 sources with high [OIII] luminosities was extensively studied using Gemini \citep{liu13a, liu13b}, and new \hst\ observations of these objects (GO-13307, PI Zakamska) were conducted in 2013 -- 2014 and are presented here. We analyze the 20 objects from programs GO-9905 and GO-13307, measuring their scattering geometry and estimating scatterer densities. Finally, the most [OIII]-luminous type 2 quasar in the \citet{reye08} catalog -- IRAS 09104+4109 -- is one of the first type 2 quasar candidates known \citep{klei88}. It was observed using the \hst\ and analyzed by \citet{hine99}, and we use some of their measurements in this paper. Thus the total sample size is 21 sources, comprised of 11 new observations and 10 archival ones. The [OIII] and infrared luminosities of our targets (as measured using {\it Wide-Field Infrared Survey Explorer}, \citealt{wrig10}) are shown in Figure \ref{pic_example} and the sources are tabulated in Table \ref{tab:sample}.

The new observations with the \hst\ -- which we discuss below in detail -- are similar in concept to the archival ones presented by \citet{zaka06}, with differences in filter selection due to differences in redshift. In the new program, each target is imaged for one orbit blueward of the 4000\AA\ break, with a typical effective wavelength near 3000\AA\ (hereafter described as rest-frame $U$-band observations) and for one orbit in the rest-frame yellow (between $V$ and $R$), with both bands carefully chosen to sample the continuum and to avoid strong forbidden emission lines. If the scattering efficiency is roughly wavelength-independent \citep{kish99}, then the spectral energy distribution of the scattered light is the same as that of an unobscured quasar -- i.e., this component is blue -- whereas in the absence of strong star formation the host galaxy is expected to be red. Therefore, by observing blueward of the 4000\AA\ break we maximize sensitivity to the scattered light and by observing redward of the break we concentrate on the stellar component of the host galaxy. In practice, both components contribute to each band, and we perform detailed analysis of the stellar component and isolate the scattered component as described in Section \ref{sec:subtract}.

In this paper we present $U$-band-based measurements of the scattered light, such as the opening angles of the scattering cones and their surface brightness profiles, and we estimate the inclination angle of cone axes to the line of sight (Section \ref{sec:model}). Furthermore, we use the measurements of the surface brightness of the scattered light to estimate the density of the scattering particles (electrons or dust), which is possible because the observed surface brightness of scattered light is proportional to the column density of scatterers along the line of sight, and we discuss the implications of our results in Section \ref{sec:discussion}. The yellow-band images are used to investigate galaxy morphology and the immediate environments of the host galaxies and are presented in the companion paper by \citet{wyle15}.

The choice of instrument is driven by the available filters, their widths and throughput, and by the requirement to avoid strong narrow emission lines such as [OII]$\lambda$3727\AA, [OIII]$\lambda\lambda$4959,5007\AA\AA, H$\beta$ and H$\alpha$. The high throughput of blue filters on the Advanced Camera for Surveys (ACS; \citealt{siri05}) determines the choice of the camera, so we use ACS F475W for objects with $z < 0.5$ and ACS F435W for the remaining two targets. In the yellow band, we again used ACS with the ramp filter FR914M, suitably placed between [OIII]$\lambda$5007\AA\ and H$\alpha$ depending on the redshift.

For each target, the ACS Wide Field Channel consists of four exposures obtained by using the default {\sc acs-wfc-dither-box} pointing pattern designed for optimal half-pixel sampling. To optimize the quality of the data products, we reprocess the data using the AstroDrizzle task in the software package DrizzlePac distributed through PyRAF. As our targets are single compact sources, we set the size of the shrunk pixels (``drops'') in the drizzle algorithm to be half of the native plate scale (0.05\arcsec), following the suggestion in the \hst\ DrizzlePac Handbook \citep{gonz12}. The adopted drizzle pattern also facilitates rejection of cosmic rays and detector artifacts. For the purpose of quality control, we have verified that statistics performed on the drizzled weight images yield a r.m.s.-to-median ratio of $\sim0.1$, satisfying the $<0.2$ requirement for balancing between resolution improvement and background noise increment due to pixel resampling, as per the \hst\ Dither Handbook \citep{koek02}. Accordingly, the final pixels of our drizzled images are resampled to 0.025\arcsec. We re-reduce archival data from GO-9905 \citep{zaka06} using the same methods.

\section{Measuring and modeling giant scattered light nebulae}
\label{sec:model}

\subsection{Host galaxy subtraction and identification of scattering cones}
\label{sec:subtract}

We make use of the rest-frame yellow observations to remove contamination by the stellar component of the host galaxy from the $U$-band. Both $U$-band and yellow-band images are a combination of the light from the stars in the host galaxy and of the scattered light, with possibility of dust extinction affecting both of these emission components. Although the yellow-band images are dominated by the stellar light, they might also contain a contribution from the scattered light of the quasar, and therefore we cannot simply subtract a scaled version of the yellow images from the $U$-band images.

Since even small offsets of a few pixels in the coordinate systems of the yellow and $U$-band images may impact the interpretation of the cleaned $U$-band images, we first carefully align the images in the two bands. We use {\tt iraf imexam} to measure the centroids of two or three stars in the field of view where the $U$-band and yellow images overlap and then use {\tt iraf imalign} to shift the $U$-band images and align them with the yellow images.

With astrometric adjustment in hand, we use {\tt galfit} \citep{peng02} to model the stellar component in the yellow-band images. We fit the two-dimensional surface brightness distributions in the yellow-band images with one or two Sersic components \citep{wyle15}. The {\tt galfit} models are our best estimates of the distribution of the stellar component, and to subtract this component from the $U$-band image we only need appropriately normalize the model.

\begin{figure*}
\centering
\includegraphics[width=\textwidth]{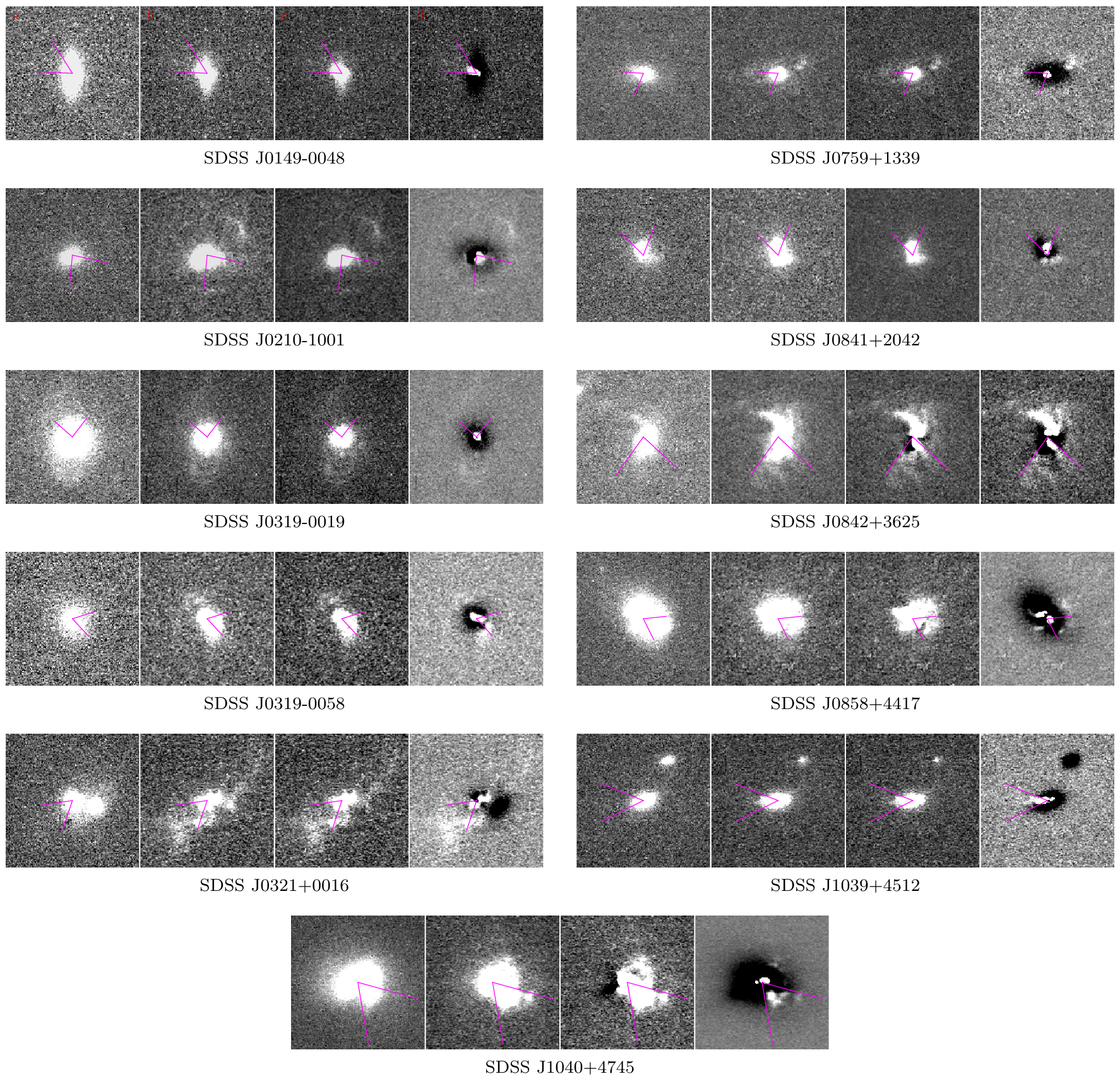}
\caption{In each strip, we show (from left to right) $4.3 \times 4.3$ arcsec$^2$ postage stamps of the (a) yellow-band \hst\ images, (b) $U$-band \hst\ images, (c) host galaxy subtracted $U$-band images and (d) extreme host galaxy subtracted $U$-band images. In (c), we subtract the scaled {\tt galfit} fit to the yellow-band image from the blue image to remove the extended stellar component, however a centrally concentrated quasi-spherical component still remains. In (d), we show extreme subtractions in which we over-subtract the stellar component to bring out the scattered light detection. Depending on redshift, the sizes of these images correspond to physical sizes of $25$ to $30$ kpc on the side. The magenta lines indicate the orientation of the forward-facing scattered light cone, and the projected angle (as shown) is used in our modeling as an upper limit on the deprojected (intrinsic) opening angle of the cone.}
\label{pic_subtract}
\end{figure*}

To this end, we measure the U-band and yellow-band fluxes within the same $1\times1$~arcsec$^{2}$ region offset from the center of the quasar by 1~arcsec and we use the ratio $f$ of $U$-band to the yellow-band fluxes within this region to scale the {\tt galfit} model derived from yellow-band images for subtraction from the $U$-band images. By performing this subtraction procedure, we assume in effect that the color of the stellar component of the host galaxy is the same as in the chosen off-center ($\sim 6$ kpc) patch. We choose this region size and offset to ensure minimal contamination by dust lanes which are strongest close to the center and we specifically avoid scattered light. Choosing a box centered on the quasar (and therefore more likely to be contaminated by strong scattered light) for {\tt galfit} normalization would lead to an overestimation of the stellar contribution to the $U$-band image. We then multiply the model yellow images by the derived scaling factor $f$ and subtract them from the $U$-band images. Almost no negative residuals result from this subtraction, which we will refer to as `optimal subtractions' hereafter. If slight negative residuals arise from this subtraction, these host galaxies typically show morphological disturbances, i.e. are undergoing or have currently undergone a merger, with dust lanes and tidal tails leading to this slight oversubtraction (see e.g. SDSS~J0842+3625 or SDSS~J1040+4745 in Figure \ref{pic_subtract}). In a successful optimal subtraction (e.g., in SDSS~J0841+2042), the scattering cone stands out nicely after the scaled host galaxy contribution is removed.

The optimal subtraction method is quite conservative in that it is optimized to subtract the contribution of the old stellar population as probed by the yellow-band images. Therefore, if there is a strong gradient of the stellar populations within the galaxy or other unaccounted for sources of $U$-band light, positive residuals emerge as a result of our subtraction in the locations in the host where stellar populations are younger. A particularly difficult host galaxy subtraction example is SDSS~J0149-0048 (Figure \ref{pic_subtract}). Even after the scaled {\tt galfit} model is subtracted from the $U$-band image, a smooth round component remains in the image. A similar component is seen in IRAS 09104+4109 by \citet{hine99}.

Aside from the morphology (featureless and centered on the quasar) and the color (bluer than the outskirts of galaxies used for {\tt galfit} normalization), we know little about this component. One possibility is that it is due to circumnuclear star formation, as suggested by \citet{hine99}. In SDSS~J1039+4512, where the optimal subtraction leaves a strong smooth component, the infrared-based measurement of star formation rate is $\sim 40$ M$_{\sun}$ yr$^{-1}$ \citep{wyle15}, so it may well host a significant unobscured young stellar population for which the optimal subtraction method is not able to account. An alternative possibility for the origin of the unsubtracted blue component -- quasar light percolating through patchy obscuration and then scattered off the interstellar medium -- is discussed in Section \ref{sec:unification}.

In some cases even with subtraction of the scaled {\tt galfit} model it is challenging to identify the scattered light component. Therefore, we also perform an `extreme' subtraction, where we subtract $6 \times f \times $ yellow-band model image from the $U$-band image. These typically yield strong negative residuals (because much of the stellar component of the host galaxy has now been over-subtracted), but often make the scattering cones stand out clearly, like in SDSS~J0149-0048 or SDSS~J1039+4512 (Figure \ref{pic_subtract}). For measurements of the scattered light geometry described in subsequent sections we use both subtractions, which provides us with an estimate of the systematic uncertainties involved.

The total flux in the $U$-band and the flux after the `optimal' stellar component subtraction are listed in Table \ref{tab:sample}. We find that the the median correction for the host galaxy contribution to the $U$-band is only 27\% and that therefore most of the flux (73\%$\pm$25\%, median and standard deviation) of the $U$-band images is due to the scattered light from the quasar. If not properly accounted for, this component can strongly bias the star formation rates of quasar host galaxies as measured from the $U$-band images.

\subsection{Scattering efficiency}

We define scattering efficiency as the fraction of the intrinsically emitted quasar radiation at $\sim 3000$\AA\ which is scattered into our line of sight. Due to the obscuration, the intrinsic amount of the $U$-band emission produced by the quasar is not known, so we estimate it using the rest-frame 12\micron\ monochromatic luminosity $\nu L_{\nu}$[12\micron] as measured from {\it WISE} data. We assume that although the observed 3000\AA\ flux is suppressed by extinction, $\nu L_{\nu}$[12\micron] is unaffected by it, so we use the latter and the average quasar spectral energy distribution from \citet{rich06} to estimate the intrinsic 3000\AA\ luminosity. The scattering efficiency is then the ratio between the actual 3000\AA\ luminosity of our objects (k-corrected as necessary using $F_{\nu}\propto \nu^{-0.44}$, \citealt{vand01}) to the estimated intrinsic value.

The median scattering efficiency in our sample is 2.3\%, with a standard deviation of 0.4 dex (Figure \ref{pic_example}, right). If 12\micron\ luminosities are in fact affected by extinction, as suggested by the red mid-infrared colors of the quasars in our sample \citep{liu13b}, then the observed 12\micron\ luminosities underestimate the true luminosities of our quasars and the measured scattering efficiencies serve as upper limits on the actual values. The scattering efficiency can be related to the geometry of the scattering regions and to the total amount of scattering mass; further discussion of scattering efficiency and the implications of the 2.3\% average value is presented in Section \ref{sec:density}.

\subsection{Estimates of extents, projected opening angles and inclination angles}
\label{sbs:iden}

We identify scattered light regions primarily by morphology in the `optimal' and `extreme' host-subtracted $U$-band images. We look for conically shaped features with apex coinciding with the center of the galaxy. Identification is aided significantly by the polarimetric observations available for approximately half of the sample \citep{zaka05, zaka06}: the polarization position angle of scattered light is expected to be orthogonal to the position angle of the major axis of the scattering cone in the plane of the sky, and indeed this relationship is borne out in our previous observations. To identify the projected axis of the cone, we look at the surface brightness distribution along annular sections taken around the galaxy center and find the peak surface brightness which corresponds to the `spine' (thickest part) of the cone. Because of the forward-scattering nature of dust (Section \ref{subsec:scamed}), if the cone axis is not exactly in the plane of the sky the observer-facing cone appears brighter and we may or may not be able to see the backward-facing cone. We find that in $\sim 14$ objects the second peak corresponding to the backward-facing counter-cone is also visible.

\begin{figure*}
\centering
\includegraphics[scale=0.3, clip=true, trim=3cm 0cm 0cm 0cm]{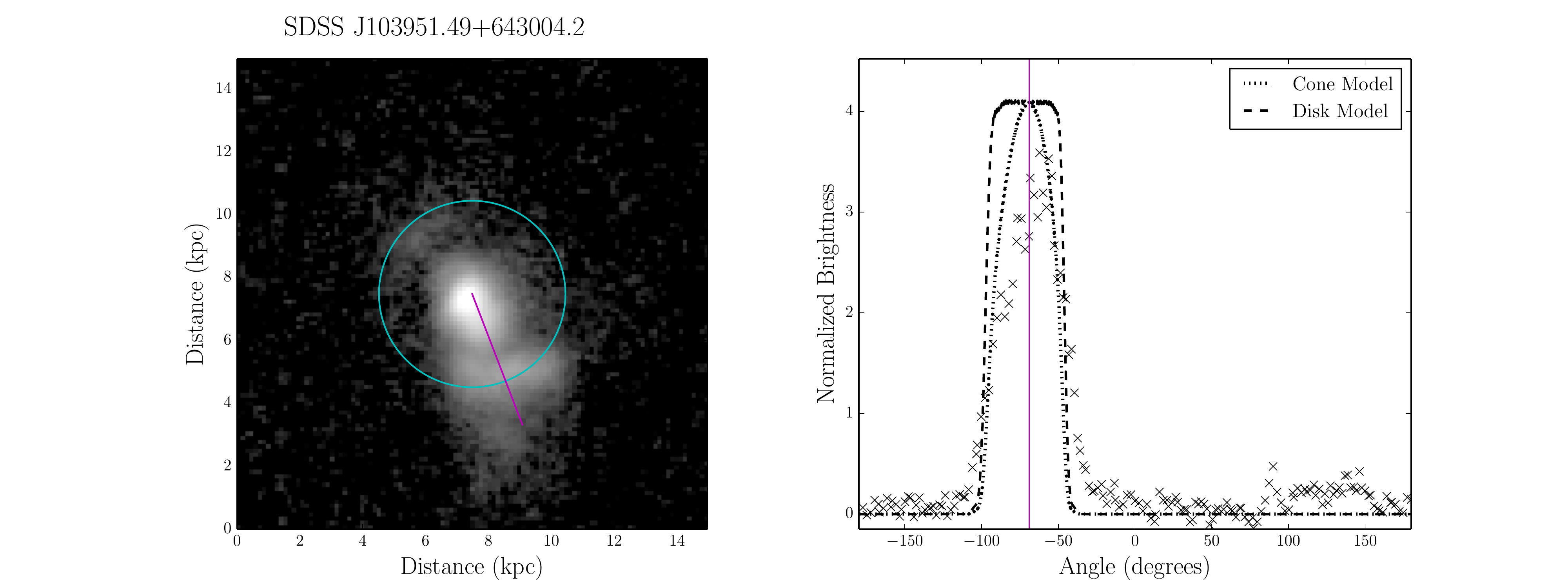} %
\includegraphics[scale=0.3, clip=true, trim=3cm 0cm 0cm 0cm]{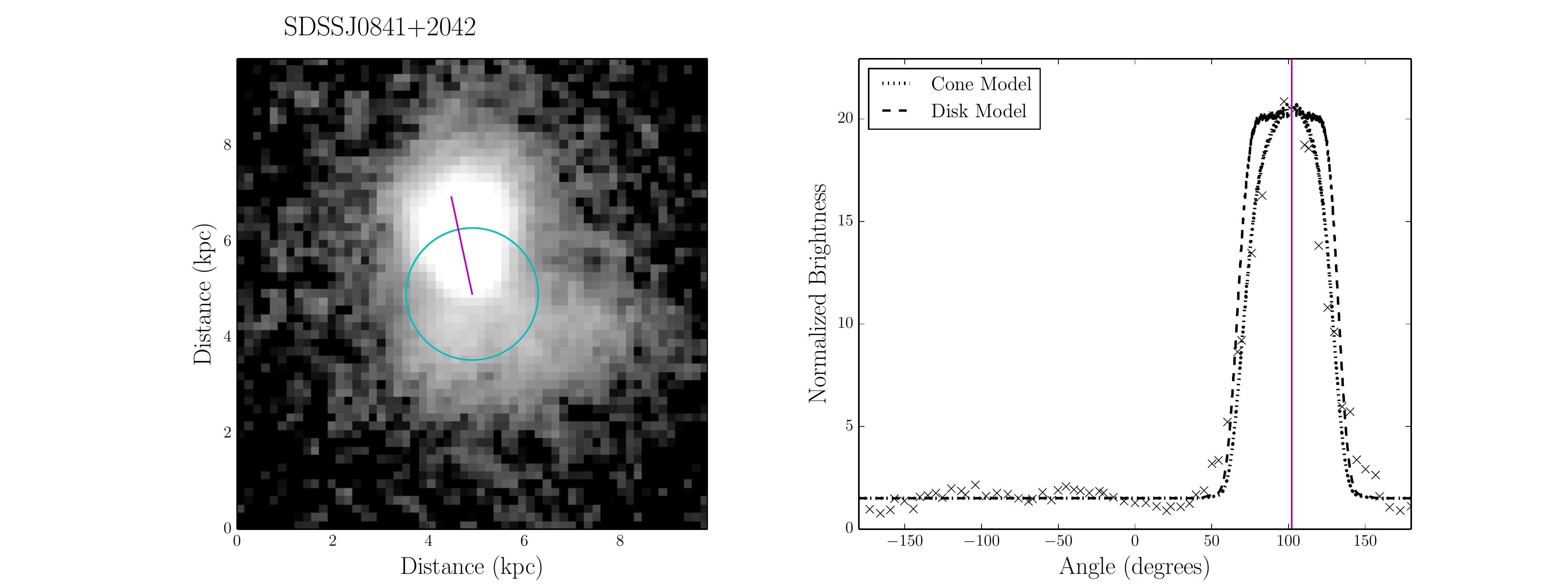}
\caption{Surface brightness profiles along a circular aperture in SDSS~J1039+6430 and in SDSS~J0841+2042. Left: optimally-subtracted $U$-band images with the location of the circular aperture (blue circle) and the direction of the spine of the scattering cone (purple line). Right: the observed surface brightness profiles (shown with crosses) are fitted with cone models (dotted lines) and disk models (dashed lines) discussed in Section \ref{sec:disk}, both convolved with the Airy function to represent the blurring by the point-spread function. Disk models show a more abrupt cut-off in surface brightness than is observed, and cone models provide a better fit to the data. Angles are counted counter-clockwise from the positive horizontal direction; counter cones are detected at $\sim 130\dg$ in SDSS~J1039+6430 and possibly from $\sim -130\dg $ to $\sim -20\dg$ in SDSS~J0841+2042 (the morphology of this component is ambiguous and it could be due to under-subtracted star formation in the nucleus of the galaxy).}
\label{pic_disk}
\end{figure*}

After identifying the directions of the cone, we trace the extents of the cones down to a limiting surface brightness $\lambda' I_{\lambda'}\simeq 3\times 10^{-15}$ erg sec$^{-1}$ cm$^{-2}$ arcsec$^{-2}$ at observed wavelength $\lambda'$. These maximal projected extents are listed in Table \ref{tab:params}. In the majority of objects the cones are detected out to at least 5 kpc from the nucleus, and in a few out to $\ga 10$ kpc.

In our subsequent calculations we assume that the apex of the scattering cone coincides with the hidden quasar, which in turn we assume is at the exact center of the stellar component of the galaxy derived from {\tt galfit} yellow-band fits. We made $<0.5$ kpc adjustments to the assumed center positions in four objects (SDSS~J0841+2042, SDSS~J1039+4512, SDSS~J1106+0357 and SDSS~J1413-0142) and a 0.9 kpc adjustment in SDSS~J0842+3625 as suggested by the position angles, morphologies and orientations of the cones.

With the center and the cone direction in hand, we can now measure the radial surface brightness profile and the lateral surface brightness profile measured along a normal to the cone's spine. The latter are taken at average distance $\sim 1.6$ kpc from the center, avoiding regions of strong surface brightness variations due to clumps. They can be used to estimate the projected half-opening angles (median 29\dg, standard deviation in the sample 8\dg) by evaluating the lateral extent of the cone in comparison to the distance from the quasar at which the lateral slice is taken. Because the cone axis is not necessarily in the plane of the sky, the opening angles as seen in projection on the plane of the sky are larger than the intrinsic ones. In the next Section, we present full conical scattering models which can be used to deproject these values and calculate the actual opening angle of the scattering cones. The most uncertain scattering cone identifications are in SDSS~J0210-1001, SDSS~J0319-0019 and SDSSJ~0858+4417. As can be seen in Figure \ref{pic_subtract}, these are also the most compact. SDSS~J0210-1001 has a luminous unsubtracted $U$-band component, and the other two objects lack the characteristic triangular shape even in the `extreme' subtractions.

Dust is strongly forward-scattering \citep{drai03}. Thus if an intrinsically symmetric bicone is not lying exactly in the plane of the sky, the cone pointed toward the observer would appear brighter than the other one, and the brightness ratio of the two can yield the inclination angle relative to the line of sight $i$. We detect unambiguous counter-cones in $\sim 8$ objects (e.g. SDSS~J0321+0016 and SDSS~J0842+3625) with a median brightness ratio of 2.5, while in $\sim 6$ sources counter-cones are tentatively detected with a median brightness ratio of 4.4 (e.g., SDSS~J0841+2042). In the remaining $\sim 6$ sources no counter-cone (e.g., SDSS~J0149-0048) is detected and we place a lower limit on the brightness ratio. Brightness ratios listed in Table \ref{tab:params} are calculated from the radial profiles taken along the spine of the brighter cone and extended to the other side of the nucleus. Assuming that the cone and the counter-cone are intrinsically symmetric, the inclination angles of the axis of the cone to the line of sight can be estimated from the brightness ratio and the phase function of dust scattering; the median estimated inclination angle is 60\dg.

\subsection{Full three-dimensional model}
\label{sbs:desc}

\begin{figure}
\centering
\includegraphics[scale=0.45]{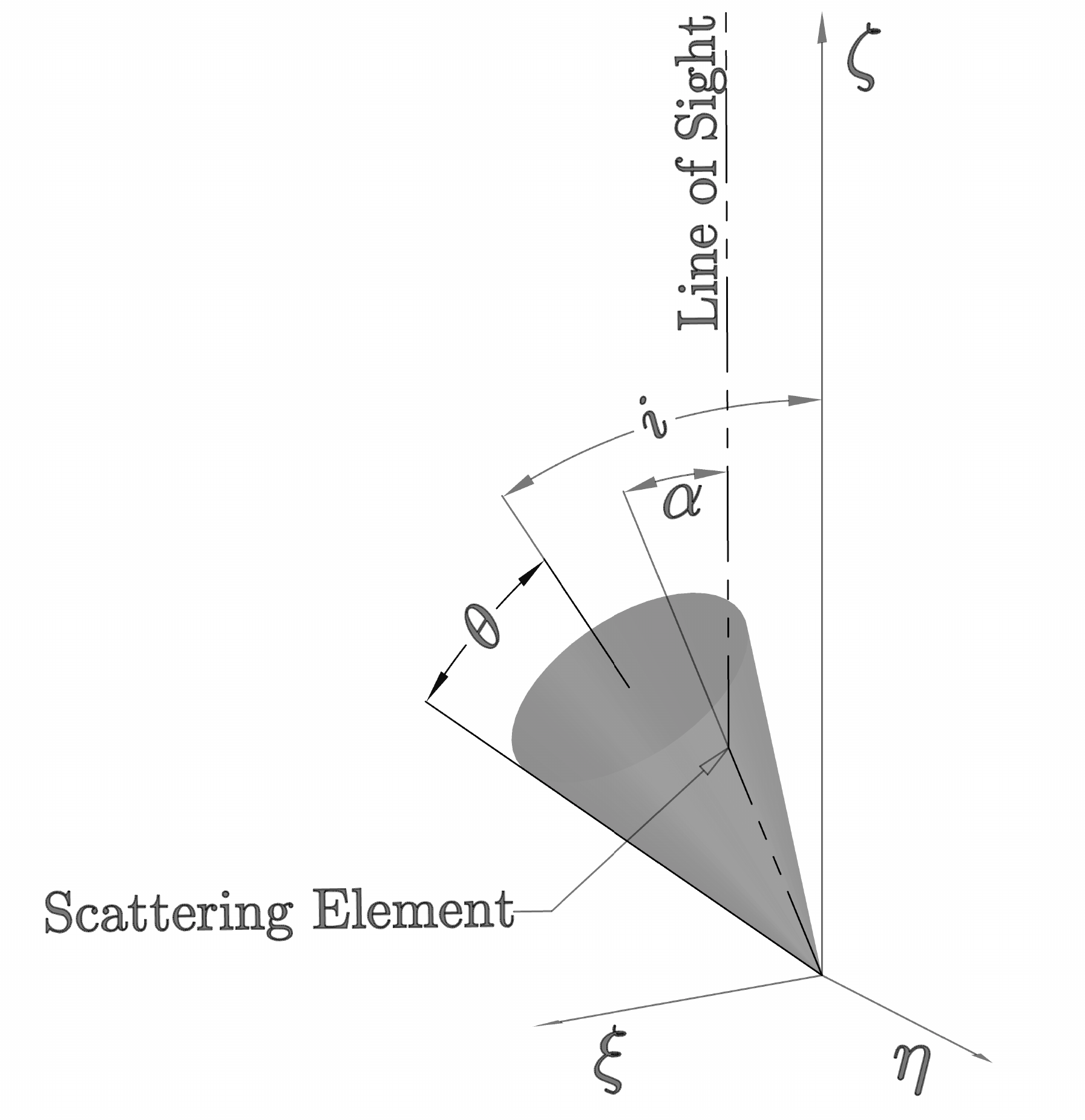}
\caption{Parameters of the scattering cone model. $\theta$ is half of the opening angle of the cone, $i$ is its inclination relative to the line of sight $\zeta$, and $\xi-\eta$ is the plane of the sky. The scattering angle $\alpha$ determines the phase function of scattering. }
\label{pic_model}
\end{figure}

We model the scattering region as a cone of half-angle $\theta$ inclined at an angle $i$ from the line of sight (Figure \ref{pic_model}), filled uniformly with scatterers (electrons or dust particles) whose number density declines as a function of distance from the radiation center $r$,
\begin{equation}
n(r)=n_0\left(\frac{r_0}{r}\right)^{2m},
\end{equation}
and the normalization is given as $n_0$ at some fixed distance $r_0$ (set equal to 1 kpc in our calculations) from the center. The observed surface brightness is then obtained by integrating the radiation scattered into our direction by every volume element of the cone along the line of sight $\zeta$:
\begin{equation}
I_{\lambda'} = \frac{(206265)^2}{(1+z)^5}\int \frac{L_{\lambda}}{4\pi r^2} n(r) \frac{{\rm d}\sigma}{{\rm d}\Omega} {\rm d}\zeta,
\label{eq_sb}
\end{equation}
where the integral ranges from the back wall of the scattering cone to its front wall. Here $L_{\lambda}$ is the intrinsic luminosity density of the quasar at rest-frame wavelength $\lambda$ (in units of erg s$^{-1}$ \AA$^{-1}$) and $I_{\lambda'}$ is the measured surface brightness (in units of erg s$^{-1}$ cm$^{-2}$ arcsec$^{-2}$ \AA$^{-1}$) at the observed wavelength $\lambda'=(1+z)\lambda$. Other parameters include $r$ -- the physical distance between the scattering volume element and the radiation center, and ${\rm d}\sigma / {\rm d}\Omega$ -- the differential cross section of scattering in cm$^2$/sr. The factor $(1+z)^5$ reflects cosmological surface brightness dimming of the luminosity density, and the factor $(206265)^2$ converts steradians into arcsec$^2$. The first crude scattering model for SDSS~J1039+6430 was presented by \citet{gree11} to estimate the density of scattering gas. Here we take into account the full geometry of scattering and appropriate scattering cross-sections for the first time.

In a partly or fully ionized medium with the standard gas-to-dust ratio, dust scattering dominates over electron scattering by a factor of $\gg 10$ at ultra-violet and optical wavelengths \citep{wein01, drai03}. Therefore, in our models we focus on dust scattering, and we present an extensive discussion of the scattering mechanism in Section \ref{subsec:scamed}. We use the Small Magellanic Cloud dust with reddening parameter $R_V \equiv A_V/E(B-V) = 2.87$ whose scattering cross sections and phase functions are given by \citet{wein01} and \citet{drai03}:
\begin{equation}
\left(\frac{d\sigma}{d\Omega}\right)_{\rm dust} = C_{\rm sca}\times p(\alpha).
\label{eq_sigmadust}
\end{equation}
Here $\alpha$ is the scattering angle between the observer's line of sight ($\zeta$-axis) and the initial radiation direction from the quasar accretion disk to the scattering event (Figure \ref{pic_model}) and $p(\alpha)$ is the phase function (angular dependence) of scattering \citep{drai03}. The wavelength dependence of total scattering cross-section $C_{\rm sca}$ for various dust compositions is tabulated by \citet{wein01}, and we take values appropriate for the rest-frame of our observations. \citet{wein01} and \citet{drai03} present scattering cross-sections normalized per hydrogen nucleus, so instead of the density of scattering dust particles we derive the density of hydrogen nuclei at 1 kpc from the quasar $n_{\rm H,0}$ under the assumption that the gas-to-dust ratio in our objects is similar to that of the Small Magellanic Cloud.

We use the Airy disk with a full width at half maximum of 0.1 arcsec as a point-spread function. This is comparable to the width of the numerically calculated ACS point-spread function obtained from \texttt{TinyTim}. A two-dimensional map of the model surface brightness is convolved with the Airy function before taking sections to obtain the model one-dimensional brightness profiles. These are then the models we fit to the radial and lateral brightness profiles taken from the data. As the first approximation, we are interested in estimating the typical densities of the interstellar medium and the radial density distributions in our galaxies. We therefore decided against a full two-dimensional image fitting procedure because in many objects the scattered light regions are very clumpy (e.g., SDSS~J0321+0016, SDSS~J1040+4745). The use of one-dimensional radial and lateral brightness profiles gives us the freedom to avoid such irregularities by, for instance, taking sections that do not pass through regions of unusually low/high density scattering medium. As we discuss below, there are other important systematic uncertainties in our derived densities, so they should be considered only as order-of-magnitude estimates.

We use the lateral and the radial surface brightness profiles to simultaneously fit eight parameters. The four physically meaningful parameters are density normalization $n_{\rm H,0}$, half-opening angle of the cone $\theta$, inclination angle of the cone axis from the line of sight $i$, and half-slope of the scatterers' density profile $m$. These parameters completely determine the geometry of the cone and the density profile and are listed in Table \ref{tab:params}. Two more parameters are background surface brightness values for the lateral and radial profiles and the remaining two parameters are the centroids of the surface brightness peaks in the one-dimensional lateral and radial profiles. In the radial profiles, our power-law analytical approximation for the density breaks down near the nucleus of the galaxy where circumnuclear obscuration can suppress scattered light, so we mask the nuclear points from the fits. In some objects we see clumping of the scattering gas; the strongest clumps are masked from the fit as well, which leads to an underestimate of the density of the scattering medium in these cases.

This fit is performed subject to several constraints. First, the opening angle of the cone must be smaller than or equal to the projected opening angle measured in the previous section (shown in magenta in Figure~\ref{pic_subtract}). Second, the inclination angle is greater than or equal to the inclination angle derived in the previous section. The rationale for using the measured inclination angle as the lower limit rather than as an actual measurement is that any dust within the galaxy would obscure the backward-facing cone more strongly, though this is a small effect: the inclination angles from model fits and from the brightness ratios agree to within 10 degrees in all but 3 objects. Third, we require that the inclination angle be greater than half of the opening angle so that we see the object as a type 2 quasar, i.e., our line of sight cannot pass within the cone. Without constraints on the angles, and especially the inclination angle from the brightness ratios of the cone to counter-cone, the inclination angle and the opening angle are quite degenerate with one another: a narrower cone pointing close to the line of sight has observed surface brightness profile similar to a wider cone closer to the plane of the sky.

\begin{figure*}
\centering
\includegraphics[scale=0.25, clip=true, trim=3cm 0cm 2cm 0cm]{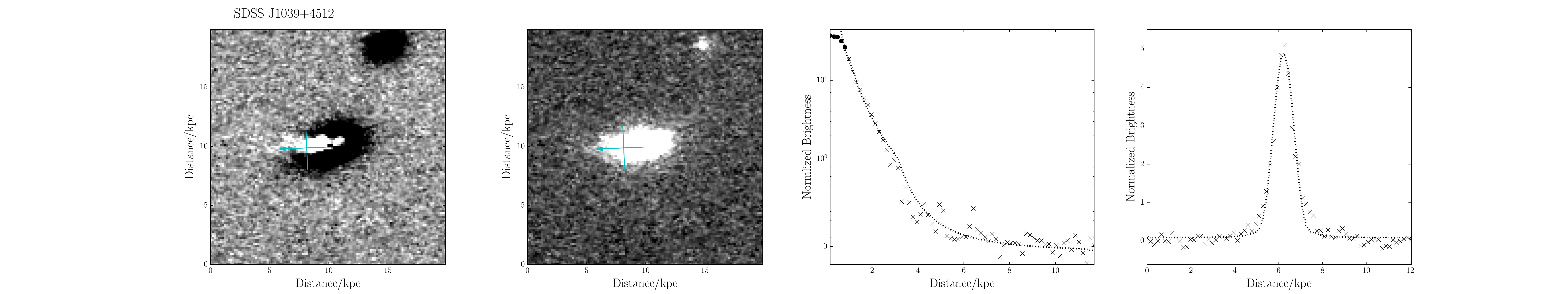} %
\includegraphics[scale=0.25, clip=true, trim=3cm 0cm 2cm 0cm]{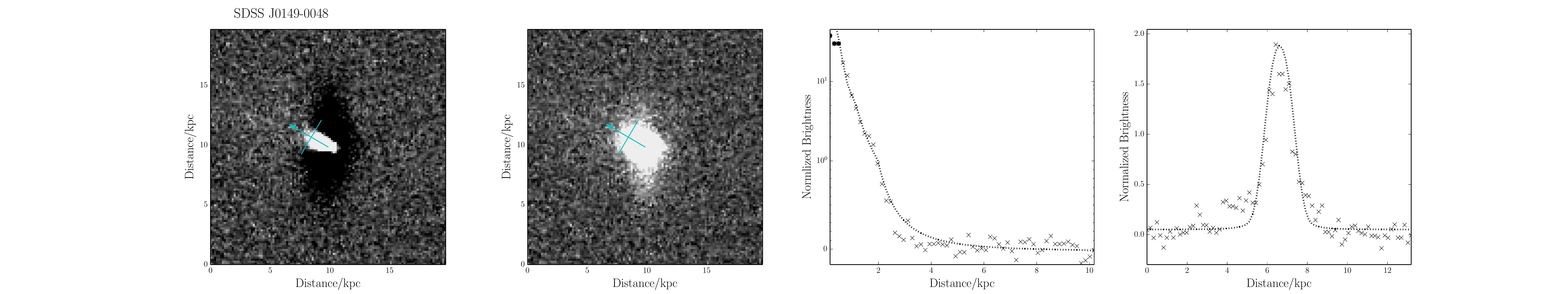} %
\includegraphics[scale=0.25, clip=true, trim=3cm 0cm 2cm 0cm]{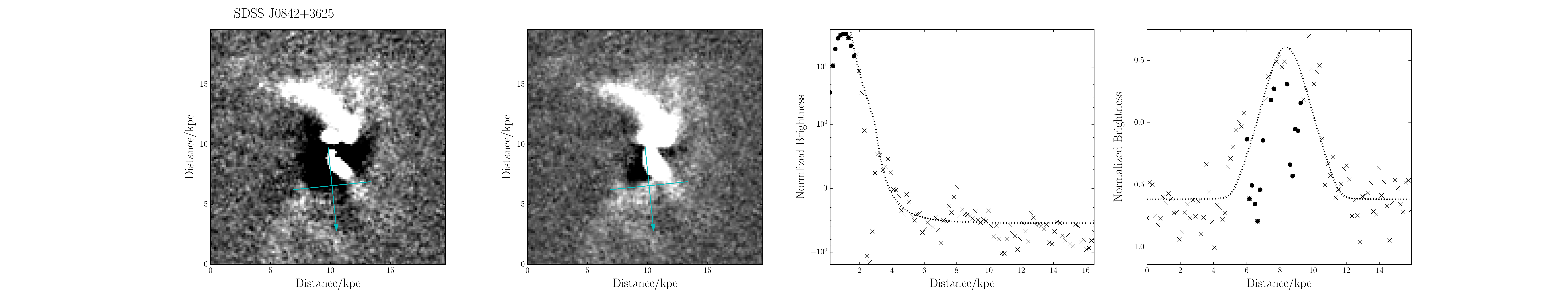} %
\includegraphics[scale=0.25, clip=true, trim=3cm 0cm 2cm 0cm]{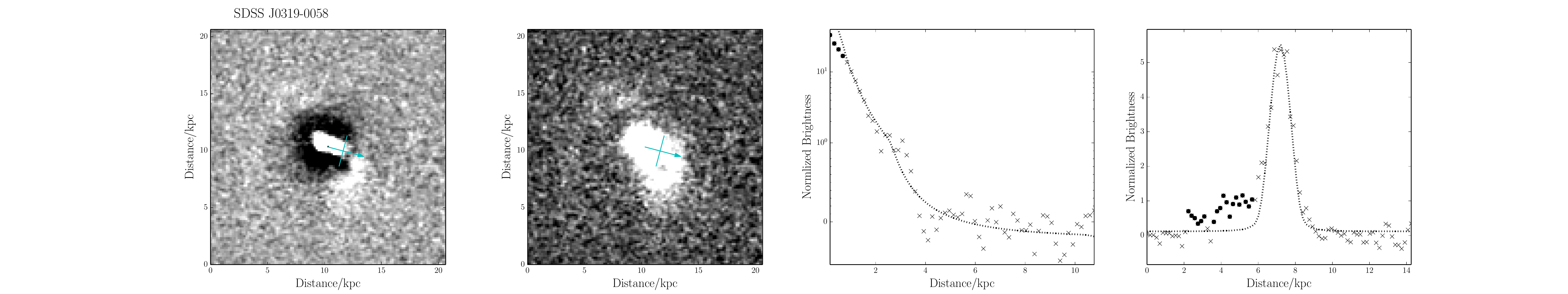} %
\includegraphics[scale=0.25, clip=true, trim=3cm 0cm 2cm 0cm]{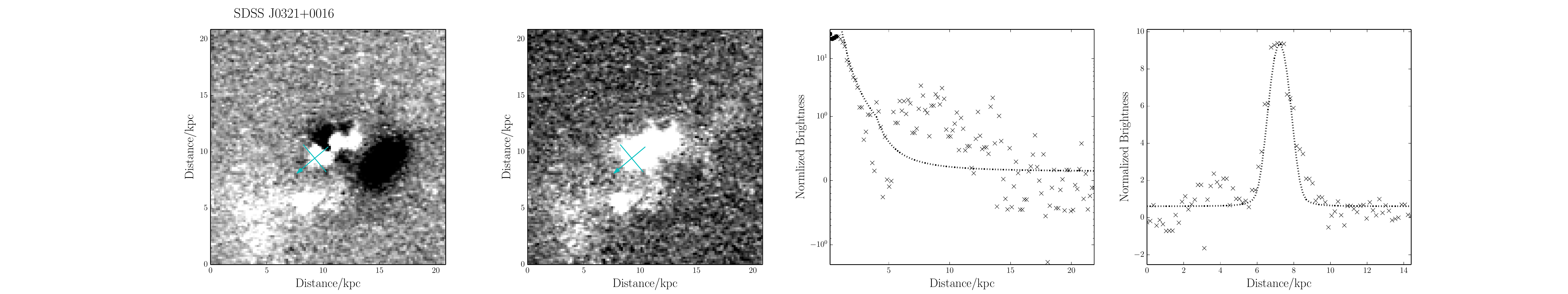} %
\caption{Examples of fitting the cone-scattering model to lateral and radial surface brightness profiles: two good-quality fits on top and three mediocre fits on the bottom. In the two left columns, we show the `extreme' subtraction of the stellar component and the `optimal subtraction' of the stellar component. The blue arrow originates at the assumed center and points along the cone spine and marks the extraction direction of the radial profile. The orthogonal line marks the location of the lateral profile. In the third column we show the observed radial profile (crosses) and the model fit (dotted line) and in the fourth column we show the same for the lateral profile. Filled points indicate data masked during the fitting process.}
\label{pic_fitting}
\end{figure*}

Fitting is done using the python \texttt{scipy.optimize} package and example fits are shown in Figure \ref{pic_fitting}. One of the sources of systematic uncertainties in our fitting procedure is due to the uncertainties in subtraction of the host galaxy. While in Table \ref{tab:params} we report the results of fitting the `optimally' subtracted images, we carry out the fitting procedure for the `extreme' subtractions as well. The fitted densities agree between the two subtractions within 0.25 dex (standard deviation), the fitted opening angles agree within 8\dg, the inclination angles within 8\dg\ and the half-slopes $m$ within 0.2. There are no significant systematic offsets between the fitted parameters in the two subtractions.

\begin{figure*}
\centering
\includegraphics[scale=0.7, clip=true, trim=0cm 10cm 0cm 0cm]{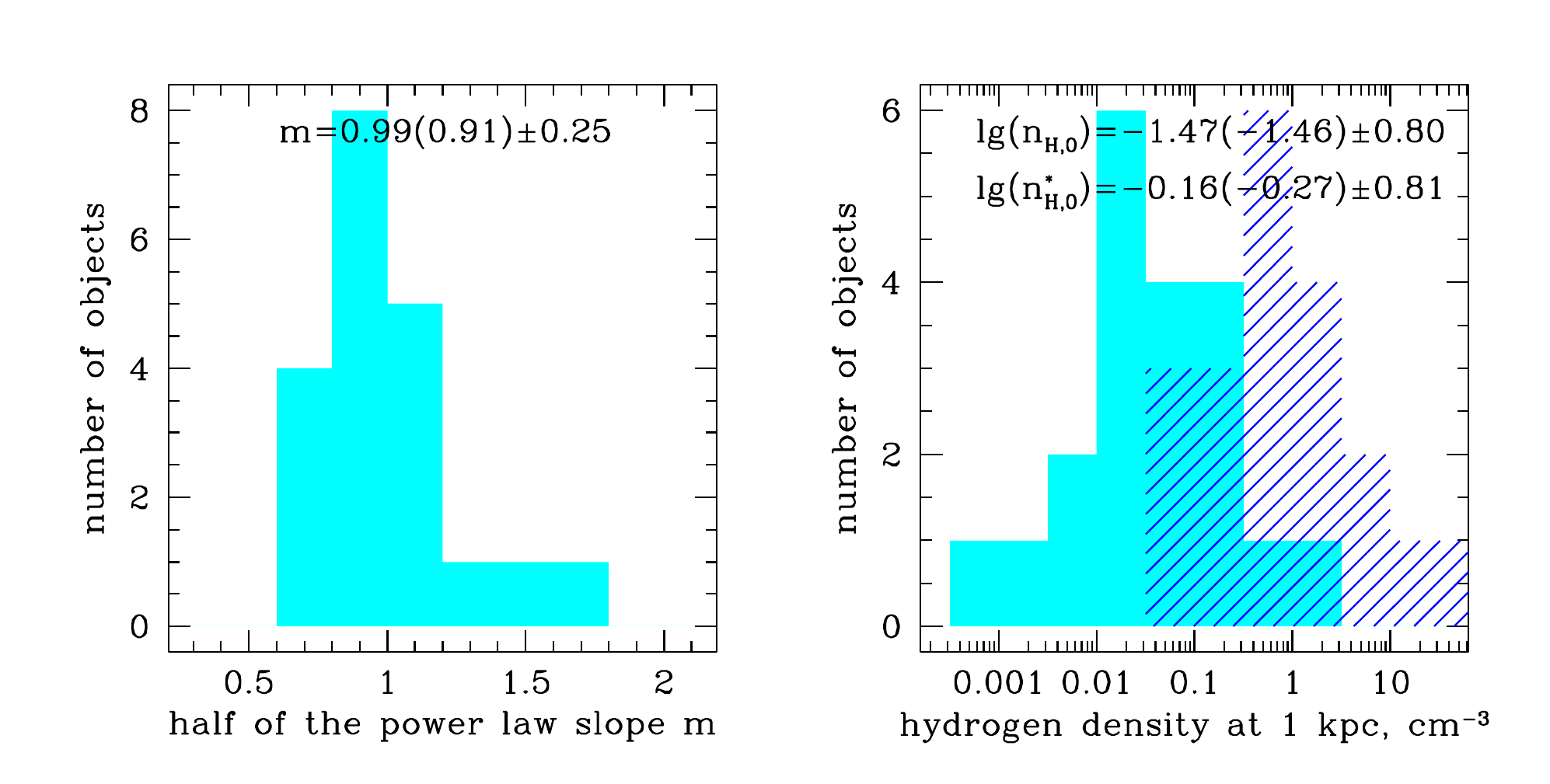}
\caption{Distribution of fitted density parameters: half of the power-law slope $m$ (left panel) and the normalization of hydrogen density at 1 kpc (right panel). We show the densities estimated using [OIII]-based bolometric luminosity $n_{\rm H,0}$ with the solid histogram and the densities estimated from the infrared luminosities $n^*_{\rm H,0}$ with the dashed lines histogram. The means (medians) and standard deviations of the distributions are shown at the top.}
\label{pic_dens}
\end{figure*}

It is clear from equation (\ref{eq_sb}) that the apparent surface brightness constrains the product of $n_{\rm H,0}L_{\lambda}$ and we need an independent estimate of the intrinsic luminosity $L_{\lambda}$ to constrain $n_{\rm H,0}$. Our default method is to use the [OIII] luminosity to derive the 2500\AA\ luminosity density from the empirical relationship known for type 1 quasars and presented by \citet{reye08}, which we then adjust to the effective rest wavelength of our observations using the average quasar spectrum $L_{\nu}\propto \nu^{-0.44}$ \citep{vand01}. The density normalizations obtained using this method are shown in Table \ref{tab:params} and in Figure \ref{pic_dens}.

An alternative method is to start from 12\micron\ luminosities (Table \ref{tab:sample}) and augment them to total infrared luminosities using average unobscured quasar spectral energy distributions from \citet{rich06}. The specific multiplicative factor that we use is 3.4. Then we assume that the infrared luminosity is due to re-radiation of the optical luminosity intercepted by the obscuring material whose covering fraction is $\cos\theta$, so that $L_{\rm opt}=L_{\rm IR}/\cos\theta$. We then use again spectral energy distributions from \citet{rich06} to connect the total optical luminosity to the monochromatic luminosity at the rest-frame of our observations ($\nu L_{\nu}[3000{\rm \AA}]\simeq L_{\rm opt}/3.$). This gives us another set of density normalizations, also shown in Table \ref{tab:params}. While the two sets of densities are well correlated with one another, the second set is significantly higher, by a factor of $\ga 10$. We suspect that the intrinsic luminosities are significantly underestimated in the second method; among other factors, $\cos\theta$ significantly overestimates the obscuration covering fraction as discussed in Section \ref{sec:unification}.

For this reason we somewhat prefer the first set of densities, but in Sec. \ref{sec:density} we use both sets of values to bracket the likely range of scattering gas masses. The comparison between the two sets clearly demonstrates the severity of systematic uncertainties in our density calculation, so we assume that our derived densities are known to no better than an order of magnitude.

\section{Results of the modeling scattered light}
\label{sec:discussion}

\subsection{Opening angles of scattering regions}
\label{sec:unification}

The mere detection of scattering cones constitutes proof of one of the key components of the classical unification model of active galactic nuclei \citep{anto85, anto93}: the objects in our sample would be seen as normal unobscured quasars if we were positioned along the lines of sight within the cones. This is demonstrated by the color of the scattered light (blue, reflecting the incident quasar spectrum) and by the spectropolarimetric observations which indicate that scattered light shows the classical broad lines characteristic of unobscured quasars \citep{zaka05}.

The average (median) and the standard deviation of the half-opening angles in our sample is $\theta = 27\dg (27\dg) \pm 9\dg$. In principle, the opening angle is related in a straightforward manner to the ratio of type 1 to type 2 quasars within the quasar population: in the case of axisymmetric toroidal obscuration, the probability of a line of sight to the observer to be located within the opening of the cone (and therefore for the object to be seen as a type 1 quasar) is $P_{\rm type1}=1-\cos\theta$, while the probability to see the object as a type 2 quasar is $P_{\rm type2}=\cos\theta$. For the observed average opening angle, the implied type 1 fraction in the population is 11\%. Although the type 1 / type 2 ratio remains somewhat controversial, this measurement is inconsistent with the typical type 1 / type 2 ratio measured from quasar demographics ($\sim 1.0$, \citealt{reye08, lawr10}) and that measured from the infrared-to-optical ratios ($\sim 2.0$, \citealt{trei08}).

One reason for the low calculated fraction of type 1 quasars in the population is that a type 2 sample selected for follow-up observations would naturally be biased toward lower opening angles, because the probability of selecting a type 2 quasar with a half-opening angle of $\theta$ increases as $\theta$ declines, so such objects would be over-represented in our sample. We construct simple models of this bias that take into account the observed $\theta$ distribution and calculate a bias-corrected type 1 fraction in the population to be $\sim 13\%$. This is still much too low by comparison with the expected $\sim 50-70\%$ value. As an extreme version of the bias, a dust-free quasar would not enter into a type 2-selected sample at all, and our bias correction procedure would not be able to account for these objects, but there are only 10\% of dust-poor quasars among optically-selected type 1 nuclei \citep{hao11}, so they do not fully resolve the discrepancy either.

To produce at 1:1 ratio of type 1 to type 2 quasars, the average half-angle of scattering cones would need to be 60\dg. We do not have a single object with $\theta$ above 40\dg. (The potential counter-cone in SDSS~J0841+2042 shown in Figure \ref{pic_disk} comes close, but it is unclear whether the observed feature is in fact a counter-cone or an under-subtracted blue stellar component.) One intriguing possibility is that the toroidal obscuration is patchy or porous, to the extent that an observer has a reasonable chance ($30-50\%$) of seeing a type 1 source even when the line of sight nominally passes through obscuration. This idea finds support in the under-subtracted blue component we see in many objects. This component was previously attributed to the possible star formation in the central regions of the galaxy (\citealt{hine99} and Section \ref{sec:subtract}), but another possible explanation is that it is also due to the scattered light, produced when quasar radiation leaks through many narrow openings in the obscuring material which cannot be identified individually but together produce the observed isotropic excess of $U$-band light.

It is becoming increasingly clear that quasar obscuration is clumpy \citep{nenk02, nenk08, scha08, zaka08, niku09, deo11, mark14}, and the probability of observing a source through obscuration as a type 1 object is determined by the covering factor of these clumps. The sizes and numbers of clumps might ultimately be constrained by spectral fitting of mid-infrared emission or by modeling of the `changing-look' active nuclei (which are interpreted as being temporarily blocked by a single cloud). It will be interesting to see whether statistics of clumps from these observations are in agreement with our scattered light data.

\subsection{Illuminated disk vs filled cone}
\label{sec:disk}

In our modeling of the scattered light regions we have assumed a cone filled with scattering particles -- or rather, we assume that the entire galaxy is filled with scatterers and the cone is produced where they happen to be illuminated by the quasar. Another possibility for explaining the observed triangular morphology of scattered light is that there is a large-scale gas disk (e.g., a disk component of the quasar's host galaxy) which is illuminated by the quasar, and the circumnuclear obscuration determines the pattern of illumination. Such component is seen in a nearby active galaxy \citep{lena15} where it is distinct from the outflowing gas component. It is important for us to make a distinction for volume-filling, potentially dynamically disturbed gas (possibly in an organized quasar-driven outflow from the galaxy) and a passively illuminated rotating galaxy disk.

Most of our objects are elliptical galaxies \citep{wyle15} without signs of large-scale galactic disks, so producing the observed scattered light components in passively illuminated disks is unlikely because we do not see a corresponding stellar component. In one of the few objects with disks (SDSS~J0149$-$0048), the stellar disk component is almost edge-on and oriented vertically in Figure \ref{pic_subtract}, whereas the scattered light cone is almost orthogonal to it (as seen in the plane of the sky). So even in this case the scattered light is not related to the disk component of the galaxy. Further analysis of the relationship between the morphologies of the stellar components and the scattered light components is presented by \citet{wyle15}.

In this section we investigate whether the observations of scattered light alone can be used to rule out the disk possibility. To this end we produce a slew of illuminated disk models with varying density profiles and inclinations and compare the resulting lateral and radial profiles between the disk and the cone cases. The biggest difference between disk and cone models is that a uniform illuminated disk in the plane of the sky would have the same surface brightness at a given distance from the quasar, whereas in the conical case the surface brightness would decline away from the cone axis. The reason for this is that at a given distance from the quasar, the disk has the same column density of scattering particles whereas the cone is thicker in the center and thinner toward the edges. Thus scattering by a gas disk should result in a sharper cutoff in surface brightness than scattering by a filled cone.

In order to exploit this difference, we take annular sections through the data and compare with conical and disk models in Figure \ref{pic_disk}. Along annular sections, data typically show surface brightness profiles with one or two peaks, depending on whether the counter-cone is detected (the two objects displayed in the figure show high contrast between the forward-scattering cone and the backward-scattering counter-cone). It is the shape of the peak that is quite different in the cone- and disk-scattering models: disk scattering predicts stubby surface brightness profiles with abrupt cut-off, so the wings of the observed surface brightness profiles are more consistent with cone-scattering models than with the disk-scattering ones. There is a hint that the wings in the observed profiles are even more extended than those in the conical models. This might indicate that the inner edge of the obscuring torus (responsible for forming the conical scattering region) is not completely opaque to escaping quasar radiation.

\subsection{The nature of the scattering medium.}
\label{subsec:scamed}

So far we have assumed that dust scattering dominates significantly over free electron scattering. Although the profiles of the scattered light regions can be equally well modeled assuming either electrons or dust, we have adopted dust scattering as our primary assumption throughout this paper, for reasons we discuss in this section.

Electron scattering is characterized by the classical Thompson cross section and phase function:
\begin{equation}
\left(\frac{{\rm d}\sigma}{{\rm d}\Omega}\right)_{\rm el} = \left(\frac{e^2}{m_e c^2}\right)^2 \frac{1+\cos^2{\alpha}}{2}.
\label{eq_sigmael}
\end{equation}
Dust scattering is more complex because in a typical astrophysical medium dust particles of many sizes are present. \citet{wein01} and \citet{drai03} use multi-wavelength observations of absorption and emission of dust in the Milky Way and in the Magellanic Clouds to derive dust size distributions and to calculate the resulting cross-section and phase function of dust scattering in these objects, normalized to the hydrogen mass (equation \ref{eq_sigmadust}).

For a normal dust-to-gas ratio typical of the Milky Way or the Magellanic Clouds, dust scattering is much more efficient than electron scattering even when hydrogen is fully ionized. Taking for example the scattering cross-sections of Small Magellanic Cloud dust at 3000\AA\ \citep{wein01}, we find that at scattering angles of $\alpha = 20^{\circ}, 40^{\circ}, 60^{\circ}, 80^{\circ}$, the minimal ratio of dust-to-electron cross section (when all hydrogen is ionized) is $\frac{{\rm d}\sigma_{\rm dust}}{{\rm d}\Omega}/\frac{{\rm d}\sigma_{\rm el}}{{\rm d}\Omega} = 220, 110, 65, 42$. This ratio is even larger for more dust-rich galaxies like the Large Magellanic Cloud or the Milky Way whose dust-to-gas ratios are $\sim 3$ and $\sim 10$ times higher than that of the Small Magellanic Cloud, respectively \citep{roma14}.

Therefore, dust scattering should dominate over electron scattering unless dust is efficiently destroyed by the radiation of the quasar or by collisions with thermal electrons (sputtering). Radiative evaporation is unlikely outside of the dust sublimation radius $r\sim 1.3 (L_{\rm bol}/10^{46}\mbox{erg s}^{-1})^{0.5}$ pc \citep{barv87}, and dust sputtering is inefficient at the typical temperatures of the narrow-line regions $T\sim 10,000$K \citep{drai79}. For these reasons, it is thought that the extended narrow-line regions of low-luminosity active galaxies are dusty, which helps explain the luminosities of narrow emission lines and the uniformity of line ratios \citep{netz93, dopi02, grov04b}, and it is possible that the conditions just outside of the broad-line regions of quasars are conducive to additional dust production \citep{elvi02}.

Despite these arguments, a closer look at dust destruction in the hosts of luminous quasars, such as the ones examined in this paper, is worthwhile. Unlike low-luminosity active galaxies, luminous quasars are capable of producing powerful galaxy-wide winds which shock the low-density phase of the interstellar medium and heat it to temperatures $\gg 10^6$ K \citep{zubo12, fauc12b, nims15}, at which dust sputtering is effective \citep{drai79} and is known to occur in hot atmospheres of galaxy clusters \citep{mcge10}. Thus dust could be destroyed in the bulk of the volume of the host galaxy, although most of dust mass should still survive in the denser phases of the interstellar medium (warm ionized and cold neutral clouds). Indeed, despite shock signatures in quasar narrow emission line regions \citep{zaka14}, line ratios are remarkably uniform and similar to those seen in low-luminosity active galaxies, suggesting that they are dusty \citep{dopi02, grov04b}. An additional complication is that the narrow-line region clouds are optically thick to ultra-violet radiation, and thus the bulk of their mass does not contribute to either dust or electron scattering. Thus the question of whether electron scattering might dominate boils down to whether the average dust-to-gas ratio of the gas exposed to the ultra-violet emission of quasars is lowered by two orders of magnitude due to sputtering.

While such calculation is outside the scope of the paper, we have several indirect arguments in favor of dust scattering from the \hst\ and polarimetric observations. Electron scattering is wavelength independent, while we see some wavelength variations of scattering efficiency in the polarized spectrum of SDSS~J1039+6430 \citep{zaka05}. Furthermore, the number of free electrons in this source is strongly constrained by the observed flux of the recombination lines and is insufficient to account for the observed scattering efficiency \citep{zaka05}. Finally, scattering by electrons in the hot low-density phase of the interstellar medium is ruled out by the lack of resulting kinematic broadening in the scattered spectrum \citep{zaka05}. These are all strong arguments in favor of dust scattering, but they require very high quality spectropolarimetric observations available for only three of the sources discussed here, SDSS~J1039+6430, SDSS~J0842+3625 \citep{zaka05} and IRAS 09104+4109 \citep{hine93, hine99}.

A larger number of objects in our sample present another signature of dust scattering -- brightness contrast between two scattering cones. Here we have to assume an axisymmetric structure for the circumnuclear obscuration, which would naturally result in two centrally symmetric scattered light bicones. If scattering is dominated by electrons, then regardless of the inclination angle of the main axis to the line of sight the two cones are expected to have the same brightness. Indeed, the cone pointed toward the observer scatters at sharp angles $\alpha$, and the cone pointed away from the observer scatters at obtuse angles $180\dg-\alpha$, but the phase function of electron scattering (eq. \ref{eq_sigmael}) is the same for these scattering directions. In contrast, dust is strongly forward-scattering \citep{drai03}, and the cone pointed toward the observer is expected to be brighter.

Out of the 21 objects in our combined sample, we find $\sim 6$ objects with just one detectable cone, $\sim 9$ objects with two cones of roughly the same brightness and $\sim 6$ objects with two cones differing in brightness by a factor of $\ga 2$. This strongly suggests that dust scattering dominates. The only alternative is electron scattering combined with dust extinction of the backward-pointing cone within the galaxy, but that requires so much dust ($N_{\rm H}=8\times 10^{21}$ cm$^{-2}$ and the Small Magellanic Cloud dust-to-gas ratio to get the median brightness ratio of $\sim 2.5$) that again dust scattering would dominate over the electron scattering in such a galaxy.

The assumption of dust scattering (as opposed to electron scattering) has critical implications for our derived values of the interstellar medium density. Because dust scattering is more efficient (by about two orders of magnitude), a smaller amount of interstellar medium is required to account for a given surface brightness of scattered light under the assumption of dust scattering than what we would calculate assuming electron scattering. We specifically choose Small Magellanic Cloud dust because \citet{hopk04} suggest that it is a good fit to the observed extinction in reddened quasars. Choosing another dust scattering curve would not noticeably affect the quality of our fits, but would affect our derived density normalization. For other types of dust like in the Large Magellanic Cloud or the Milky Way, the dust-to-gas ratio is a factor of $3-10$ times higher than for the Small Magellanic Cloud, so using these curves to fit the observed scattered surface brightness we would derive densities smaller by that factor. Self-absorption within the scattering cones is negligible; the optical depth to dust extinction within the cones is $\tau \sim n_{\rm H,0}\zeta C_{\rm ext}=0.01$ for our median $n_{\rm H,0}$ (Figure \ref{pic_dens}), $\zeta=1$ kpc and $C_{\rm ext}=1.1\times 10^{-22}$ cm$^2$.

In principle, polarimetric measurements can help distinguish between electron and dust scattering because polarization due to electron scattering can reach 100\% (when light is scattered at $\alpha=90\dg$), whereas the theoretical limit for dust-induced polarization is much lower ($\sim 20\%$ at $\alpha\sim 90\dg$ at 3000\AA, \citealt{drai03}). In practice, many factors act to lower the observed polarization fraction to typical values of a few per cent. Deviations of the inclination of the scattering cone away from the optimal directions lead to steep decline of the polarization fraction, wide opening angles of scattering region lead to geometric cancellation of the polarized signal (a centrally symmetric scattering nebula has zero net polarization), and most importantly the unpolarized light from the host galaxy dilutes the weak polarized signal.

The two objects with the highest observed levels of polarization in our sample are SDSS~J1039+6430 and SDSS~J0842+3625, both with $P=16.5\%$ at 3000\AA\ and likely negligible host galaxy dilution at this wavelength \citep{zaka05}. Taking the geometric parameters of our best-fit scattering cones in these two objects, we derive model polarization fractions of $\sim 5\%$ and $\sim 19\%$, correspondingly. The scattering cone in SDSS~J0842+3625 is in the plane of the sky, resulting in a polarized fraction which is close to the theoretical maximum for dust polarization, and the model value is consistent with the observed one. The scattering cone in SDSS~J1039+6430 is forward-scattering at $\sim 40\dg$, which suppresses polarization, and the resulting model polarization is inconsistent with the observed value. While electron scattering could easily bring the model value up to the levels consistent with observations, this object presents the strongest multi-wavelength case for dust scattering in our sample \citep{zaka05} and its inclination angle is well-determined from the detection of the counter cone (Figure \ref{pic_disk}). The only way to resolve this inconsistency is to postulate that the dust size distribution in this object is not well described by the distributions considered by \citet{drai03} for the Milky Way and the Magellanic Clouds which all produce $<20\%$ polarization at 3000\AA\ even in the optimal geometry.

\subsection{Implications of the density distribution estimates}
\label{sec:density}

The 11 objects with new \hst\ observations were studied using Gemini integral field spectroscopy by \citet{liu13a, liu13b} who uncovered galaxy-wide ionized gas outflows in these objects. If the gas detected within the scattering regions is outflowing with the typical velocities seen in the spectroscopic observations, then we can calculate the mass outflow rate: $\dot{M}(r) = \Omega r^2 n_{\rm H}(r) m_{\rm H} v(r) / X$. Here $n_{\rm H}(r)$ is the density of the interstellar medium that we obtain from our scattered light measurements, $m_{\rm H}$ is the mass of a hydrogen atom, $X\simeq 0.7$ is the cosmic hydrogen fraction by mass, and $\Omega$ is the solid angle coverage of the outflowing gas. The outflow velocity $v(r)$ can be measured in integral-field unit observations of quasar-driven winds, and in observations to date it appears almost constant as a function of distance \citep{liu13b, harr14}, with typical magnitude $v\simeq 800$ km s$^{-1}$.

While we have no direct measurement of the velocity of the gas which is responsible for the scattered light seen in the \hst\ images, we have some evidence that the scattering medium is co-spatial with the kinematically identified wind. \citet{wyle15} compared the morphologies of the $U$-band images, yellow-band images and the kinematic maps from \citet{liu13b} for the same 11 objects analyzed here. They find that the scattering cones are aligned with the direction of the velocity gradient, as expected if both the scattering cones and the photo-ionization pattern of the emission-line gas are determined by the same illumination geometry.

If only the illuminated gas is outflowing, then our estimates of the opening angles suggest $\Omega \simeq 0.1$, whereas the 1:1 obscured-to-unobscured ratio found in quasar demographic studies \citep{lawr10} suggests $\Omega \simeq 2\pi$ -- half of the sky, as seen from the quasar (this interesting discrepancy is discussed in Sec. \ref{sec:unification}). However, the circumnuclear material does not necessarily collimate the outflow toward the unobscured directions. \citep{wagn13} show that dense circumnuclear clouds disperse and deflect AGN outflows. The clouds receive most of the momentum of the outflowing gas which propagates through the low-density channels of least resistance between the dense clouds. As a result of this interaction, any directionality of the original wind vanishes on larger scales, with the outflow proceeding in all directions and curving around dense clouds and larger obstacles. In this case, the outflow would have a covering solid angle of $\Omega \simeq 4\pi$, but only part of it would be illuminated by the quasar and be detectable as scattered light, resulting in a bi-conical appearance.

The estimate for $n_{\rm H}(r)$ is a direct result of the observations presented in this paper. From our scattered light fits, we find that the median density at $r_0=1$ kpc from the quasar is $n_{\rm H,0}\simeq 0.04$ cm$^{-3}$ and that the median power-law profile of the density distribution is close to $r^{-2}$. Assuming volume-filling geometry for the outflow suggested by the scattered light observations, the estimated density slopes imply that $\dot{M}(r)$ is nearly constant as a function $r$, so that the outflowing mass does not concentrate at any one distance, which is consistent with a steady-state process of gas removal. Supplying now all the estimates into the equation for mass outflow rate, we find
\begin{equation}
\dot{M}\simeq 14.5 \frac{M_{\odot}}{\rm year}\times \left(\frac{\Omega}{4\pi}\right)\left(\frac{n_{\rm H,0}}{0.04{\rm cm}^{-3}}\right)\left(\frac{v}{800{\rm km\, s^{-1}}}\right).
\label{eq_mass}
\end{equation}
These outflow rates are much smaller than those derived from the ionized gas observations \citep{liu13b} -- $\dot{M}_{\rm ionized}\simeq 1000 M_{\odot}$ yr$^{-1}$.

The outflow mass rate is an important value for understanding what role quasar-driven winds might play in the evolution of their host galaxies, and unfortunately has proven to be difficult to obtain. The mass measurement from \citet{liu13b} is subject to many possible uncertainties, such as the electron density in the warm ionized clouds at 7 kpc from the quasar which is unconstrained from the current ionized gas observations. In the $\dot{M}_{\rm ionized}$ estimate above, we have assumed radiation pressure confinement for the narrow-line gas in agreement with \citet{dopi02} and \citet{ster15}. The chief reason that the ionized gas outflow rate derived by \citet{liu13b} is so high is that it is measured at the location where the ionized gas clouds transition from being optically-thick to ionizing radiation to being optically-thin (from ionization-bounded to matter-bounded), which was deduced from the observations of \citet{liu13a} of the increase in the HeII$\lambda$4686\AA/H$\beta$ ratio at this distance. In other words, over a small range of distances from the quasar (at $\sim 7$ kpc) the ionization conditions in these clouds are such that they are fully visible to the observer and it is unsurprising that one would derive the maximal mass rate at this distance.

At distances $r \ga 7$ kpc the gas is ionized to higher ionization levels and is invisible in optical transitions. At distances $r \la 7$ kpc most relevant for the scattered light observations only a small fraction of the volume produces observable emission lines. In the same warm ionized phase, only a small fraction of the volume of the narrow-line clouds is transparent to the ultra-violet emission. We suggest that the discrepancy between the outflow rates derived from scattered light observations and those derived from the ionized gas observations is due to the fact that most of the outflowing gas is in the form of dense clouds, and only a small fraction of the clouds' mass ($\sim 1.5\%$ at 1 kpc from the quasar, as suggested by eq. \ref{eq_mass}) participates in the scattering.

Can we construct a crude model of a clumpy narrow-line region which would be consistent with both the new scattered light observations and the previous emission-line observations \citep{liu13a, liu13b}? Assuming that clouds are radiation-pressure confined \citep{dopi02, ster15} and that they are propagating outward in a mass-conserving fashion, we find that their column density depends on the distance from the quasar as $N_{\rm H}\propto r^{-4/3}$. Clouds transition from being ionization-bounded to being matter-bounded at hydrogen column density $N_{\rm H}=10^{21}$ cm$^{-2}$ \citep{ster15}. If this transition happens at 7 kpc \citep{liu13b}, then at 1 kpc from the quasar -- the typical distance probed by our scattered light observations -- the column density of narrow-line clouds is $N_{\rm H}\sim 13\times 10^{21}$ cm$^{-2}$. Ultra-violet photons can only penetrate through and escape from a narrow layer on the surface with optical depth $\tau$, with column density $N_{\rm H, scattered}\simeq 1.8\times 10^{21}(\tau/0.2)$ cm$^{-2}$, where we have used the appropriate extinction cross-section from the Small Magellanic Cloud dust opacity curve \citep{wein01}. This suggests that at 1 kpc $10-15$\% of the mass of the narrow-line clouds should be visible in scattered light observations.

Therefore, clumping of the narrow-line region explains some (though not all) of the tension between the mass estimates produced by the two different methods. Possibly, the densities $n_{H,0}$ derived from scattered light observations are too low. As we discussed in Sec. \ref{sbs:desc}, the derived densities are degenerate with the assumed intrinsic luminosity at 3000\AA. The upper limits on the densities $n_{H,0}^*$ derived from infrared luminosities are an order of magnitude higher; using these values would result in an estimated mass outflow rate of 200 $M_{\odot}$ yr$^{-1}$, which in combination with clumping would be consistent with ionized gas observations. Another source of uncertainty in our estimates is the assumption of the single matter-bounded transition boundary. It is much more likely that a wide range of cloud sizes is present in the narrow-line region and they transition into the matter-bounded regime at different distances. We are currently developing such models and are aiming to include them in our future analyses of quasar-driven winds.

Furthermore, most of the outflow mass could be in the form of neutral or even molecular gas \citep{morg05, feru10, veil13a, cico14, sun14}. This phase would be invisible to optical emission line observations and only the thin outer layers of such clouds would participate in scattering because the molecular gas presumably would be concentrated in very dense individually optically thick clouds. Neither ionized gas observations nor scattered light observations would capture this component of the outflows.

We can now use the derived density profiles $n(r)\propto r^{-2}$ to relate the scattering efficiency $\varepsilon=\int{\rm d}V\left({\rm d}\sigma/{\rm d}\Omega\right)n(r)/r^2$ \citep{zaka05} to the total mass of gas in the galaxy out to $r_{\rm max}$ from the center:
\begin{equation}
\varepsilon=\left\langle\frac{{\rm d}\sigma}{{\rm d}\Omega}\right\rangle \frac{P_{\rm type1}M_{\rm H}}{m_{\rm H}r_{\rm max}r_{\rm min}}.
\end{equation}
Scattering is dominated by small distances from the quasar ($r_{\rm min}$) where the quasar radiation is least diluted. To make an estimate of the total hydrogen mass, we adopt ${\rm d}\sigma/{\rm d}\Omega=3\times 10^{-24}$ cm$^2$/sr, $r_{\rm max}=10$ kpc, $r_{\rm min}=0.1$ kpc and $P_{\rm type1}=0.5$ to find $M_{\rm H}=10^8M_{\odot}$. This is the lower limit on the actual gas mass since much of the mass might be invisible in scattered light as discussed above. Furthermore, this gas is not concentrated in a disk but rather is distributed through the volume of the host galaxies. The combination of moderately high gas mass and the spatial distribution of this gas makes the host galaxies of luminous obscured quasars highly unusual among early-type galaxies \citep{serr12}, supporting the hypothesis that the gas we see in scattered light is associated with quasar outflows also seen in kinematic measurements \citep{wyle15}.

To summarize, the picture we propose on the basis of the scattered light observations involves many small clouds filling the host galaxy. When the clouds are illuminated by the quasar, they are partly photo-ionized and produce the observed optical emission lines, while their surfaces contribute to the observed scattered light. An alternative geometry often discussed in the context of galactic winds is that of overpressured bubbles \citep{macl89, gree12, harr15} which expand into the interstellar medium and plow a shell of gas off to the sides and into the intergalactic space. Such shells may be responsible for the discrete-velocity features seen in absorption-line observations of quasar-driven outflows, and this geometry is often assumed in modeling these data \citep{arav08, moe09, borg12}. 

In our sample SDSS~J0319-0019 shows potential bubble walls in the emission line observations \citep{liu13b, wyle15}, and in the \hst\ image we see a circular shell-like feature whose left wall (as seen in Figure \ref{pic_subtract}) is co-spatial with one of them. As mentioned in Sec. \ref{sbs:iden}, the putative scattered component near the nucleus is compact and does not follow the regular triangular morphology. Although the pressure inside the bubble may be very high, the density of the bubble medium can be low as it is shock-heated to high temperatures. Therefore, in the case of an illuminated bubble we might preferentially see only the bubble walls both in the ionized gas and in the scattered light observations, in qualitative agreement with our observations of SDSS~J0319-0019. A weaker bubble candidate is SDSS~J0210-1001. It, too, has a shell-like feature in the $U$-band image (Figure \ref{pic_subtract}). This feature corresponds to a region of very low velocity dispersion, as discussed by \citet{liu13b} and therefore could be attributed to an illuminated dwarf galaxy companion or tidal debris. 

The observed morphology of the scattered light depends both on the underlying density distribution of the gas and on the illumination pattern. For objects which are undergoing mergers, the scattered light morphology is affected by the merger (e.g., SDSS~J0842+3625, SDSS~J0858+4417). For objects with candidate wind-driven bubbles (SDSS~J0319-0019 and SDSS~J0210-1001), the scattered light morphology is determined by the presence of the bubbles. Therefore, modeling our scattered light regions as cones filled with scatterers of declining density is only an approximation. Figures \ref{pic_disk} and \ref{pic_fitting} support the use of this approximation for estimating the surface brightness distribution of scattered light in most of the sample and indicate that this approximation is preferable to some other geometries (e.g., thin disk). 

\section{Conclusions}
\label{sec:conclusions}

In this paper we investigate \hst\ images of 20 powerful obscured quasars selected based on the [OIII] emission line luminosities. The images probe continuum emission at $\sim 3000$\AA\ in the rest frame of our targets and are sensitive to scattered light: emission from the quasar that is reflected off the interstellar medium in the host galaxy toward the observer. The spectrum of this component is expected to be similar to that of the underlying quasar and is thus much bluer than normal stellar populations, allowing us to disentangle the light from the host galaxy and the scattered light using observations at rest-frame $\sim 6000$\AA\ which predominantly probe the stellar component.

We detect luminous, extended scattered light regions in most cases. Down to a limiting surface brightness of $\lambda' I_{\lambda'}\simeq 3\times 10^{-15}$ erg sec$^{-1}$ cm$^{-2}$ arcsec$^{-2}$ at observed wavelength $\lambda'\simeq 4500$\AA, scattered light can be traced out to $\ga 5$ kpc from the nucleus in 17 objects and out to $\ga 10$ kpc from the nucleus in 3 objects. While signatures of quasar-driven outflows are now detected out to several kpc via a variety of methods (e.g., \citealt{nesv08, feru10, borg12, gree12}), only a handful of giant scattering nebulae of comparable extents had been seen before \citep{hine93, hine99, schm07}. As these scattering regions predominantly have conical / triangular appearance, they can be identified based on morphology alone. Furthermore, the orientation of the scattered light regions is in excellent agreement with ground-based polarimetric measurements available for half of this sample \citep{zaka05, zaka06}, in that the measured polarization position angles are orthogonal to the axes of the scattering cones as seen in the plane of the sky. The mere detection of conical scattered light regions in these sources implies that these sources would be seen as normal blue quasars along some lines of sight, in agreement with the foundational principle of the geometric unification model of the active galactic nuclei \citep{anto93}.

We estimate that 2.3\% of the intrinsic luminosity of the quasar is scattered off the interstellar medium and reaches the observer. Furthermore, at 3000\AA\ scattered light constitutes about 73\% of the total emission. This fraction includes both the conically shaped bright scattered-light regions and the faint quasi-spherical component which remains after subtraction of scaled yellow-band stellar models. The origin of this component is not fully understood; it may be due to star formation in the nucleus of the host galaxy, but it can also be due to scattered light produced when quasar radiation percolates through patchy obscuring material.

Failing to correct for the scattered light component in observations of hosts of luminous quasars would lead to a dramatic overestimate of the star formation rates in quasar host galaxies. The median apparent 3000\AA\ luminosity of the objects in our sample is $\nu L_{\nu}$[3000\AA]$=10^{43.88}$ erg s$^{-1}$. If we did not know that most of this luminosity is due to scattered light, we would have derived the rate of unobscured star formation of $\sim 12 M_{\odot}$ yr$^{-1}$ \citep{igle04}, having corrected the apparent luminosity from 3000\AA\ to 2000\AA\ using $f_{\lambda}\propto \lambda^{-2}$ \citep{meur99}. But because most of the ultra-violet emission is due to scattered light, the actual rates of unobscured star formation are closer to $\sim 3 M_{\odot}$ yr$^{-1}$.

Our results demonstrate the key difficulty of using quasar hosts' colors and luminosities for deriving star formation rates and stellar masses, as is commonly done for type 1 (unobscured) quasars \citep{sanc04, schr08}. Extended scattered light is expected to be present in quasars of both types, though it is difficult to say how much of a contribution it makes to the extended emission in type 1s \citep{youn09}. On the one hand, in type 1s, which face the observer, one might expect that the scattered light regions would appear more compact because of projection effects and thus would be absorbed into the point-spread function component associated with the quasar itself, in which case scattered light would not present much of a problem for the derived host values. On the other hand, because of the forward-scattering nature of dust particles one might expect the scattering efficiencies to be higher in type 1s than in type 2s, which would make scattering effects stronger in type 1s. Morphological identification of scattered light regions would be next-to-impossible in type 1 sources, both because of the bright quasar and because the scattering cone is facing the observer and is thus lacking the characteristic triangular shape we see in type 2s. Stellar population decomposition of off-nuclear spectra which take into account the possibility of scattered light may be a more reliable procedure for calculating the star formation rates of quasar hosts \citep{liu09, cana13}.

Two cones are detected in 14 of the objects, pointing in roughly opposite directions from the nucleus. This is consistent with axisymmetric toroidal obscuration postulated by the classic geometric unification model. The brightness ratios of the two cones and the absence of a second cone in the 6 remaining sources strongly suggest that dust (which is more efficient in the forward direction than backwards) is responsible for scattering. By comparing our models of dust-scattered cones with observations, we find tentative evidence that dust in quasar hosts is a more efficient polarizer at 3000\AA\ than dust in Magellanic Clouds and the Milky Way \citep{drai03}. Models with dust concentrated in a disk are inconsistent with observations; volume-filled cones produce better fits.

The measured opening angles of scattering cones allow us to calculate the type 1 / type 2 ratio from the probability that the observers' line of sight lies within the scattering cones, assuming that those correspond exactly to the opening angles of the obscuring material. The resulting type 1 quasar fraction in the population is only $\sim 13\%$, inconsistent with many studies suggesting that it is $\ga 50\%$ at these redshifts and luminosities. This is an interesting discrepancy which is not easily discounted as due to problematic opening angle measurements, as we do not see a single scattering cone with the half opening angle of 60\dg\ expected for a 1:1 ratio of type 1 to type 2 quasars. We suggest that the actual type 1 fraction can be higher than that derived from the opening angles if the obscuring material is patchy and if $30-50\%$ of the lines of sight going through it nonetheless result in a type 1 appearance. This hypothesis would not only reconcile the measured opening angles with the studies of quasar demographics, but would explain the excess spherically-symmetric $U$-band emission left over after stellar subtraction and the morphology of the emission-line nebulae which are more isotropic than would be expected in conical illumination \citep{liu13a, liu13b}.

Detection of extended scattered light offers a unique opportunity to estimate some properties of the interstellar medium of the host galaxies of very luminous quasars. From modeling scattered light regions we find that the typical profile of the density of the interstellar medium responsible for the observed scattered light is $n_{\rm H}(r)\simeq 0.04-0.5$ cm$^{-3} \times (r/1{\rm kpc})^{-2}$. There are large systematic uncertainties in the normalization of this density and it should be regarded as an order-of-magnitude estimate. The slope of this density profile is consistent with that established in a steady-state, constant velocity outflow. If this gas participates in an outflow with typical velocities $v \simeq 800$ km s$^{-1}$ suggested by the emission-line observations \citep{liu13b}, then the mass outflow rate of the interstellar medium seen in scattered light is $\dot{M}\sim 15-200 M_{\odot}$ yr$^{-1}$. This is only $1.5-20$\% of the previously measured outflow rates of the narrow-line-emitting gas. We suggest that if the outflow is primarily in the form of dense clouds (either in the warm ionized phase or in the cold neutral or molecular phase) then these clouds are likely to be optically thick to ultra-violet emission and the bulk of their mass is not participating in scattering, so only thin outer shells of the clouds facing the quasar would contribute to scattered light.

\acknowledgments

The authors are grateful to Julian Krolik and the anonymous referee for useful discussions. Based on observations associated with programs GO-13307 and GO-9905 made with the NASA/ESA {\it Hubble Space Telescope}, obtained at the Space Telescope Science Institute (STScI), which is operated by the Association of Universities for Research in Astronomy, Inc., under NASA contract NAS 5-26555. Support for program GO-13307 was provided by NASA through grant HST-GO-13307.01-A from the STScI. G.O. acknowledges support by the Provost's Undergraduate Research Award at Johns Hopkins University. D.W. acknowledges support by Akbari-Mack Postdoctoral Fellowship.

\bibliographystyle{apj}
\bibliography{master}

\clearpage
\begin{deluxetable}{l|l|l|l|l|l|l}
\tablecaption{\hst\ observations of quasar scattered light nebulae\label{tab:sample}}
\tablehead{ID  & $z$ & $L$[OIII] & $\lambda_{\rm eff}$ & $F_{\nu}$ & subtracted $F_{\nu}$ & $L$[12\micron]}
\startdata
SDSS J014932.53$-$004803.7 & 0.567 & 42.87 & 3031 & 8.7 & 7.2 & 45.21 \\ 
SDSS J021047.01$-$100152.9 & 0.540 & 43.48 & 3084 & 13.8 & 11.6 & 44.87 \\ 
SDSS J031909.61$-$001916.7 & 0.635 & 42.74 & 2905 & 17.0 & 10.1 & 45.01 \\ 
SDSS J031950.54$-$005850.6 & 0.626 & 42.96 & 2921 & 5.3 & 4.4 & 44.89 \\ 
SDSS J032144.11$+$001638.2 & 0.643 & 43.10 & 2891 & 4.1 & 3.4 & 45.10 \\ 
SDSS J075944.64$+$133945.8 & 0.649 & 43.38 & 2881 & 6.7 & 5.6 & 45.47 \\ 
SDSS J084130.78$+$204220.5 & 0.641 & 43.31 & 2895 & 7.3 & 6.4 & 45.09 \\ 
SDSS J084234.94$+$362503.1 & 0.561 & 43.56 & 3043 & 15.6 & 5.8 & 45.06 \\ 
SDSS J085829.59$+$441734.7 & 0.454 & 43.30 & 2992 & 44.4 & 32.4 & 45.80 \\ 
SDSS J103927.19$+$451215.4 & 0.579 & 43.29 & 3008 & 9.1 & 7.9 & 45.30 \\ 
SDSS J104014.43$+$474554.8 & 0.486 & 43.52 & 2927 & 60.6 & 43.4 & 45.48 \\ 
\hline
SDSS J012341.47$+$004435.9 & 0.399 & 42.71 & 3109 & 9.1 & 3.6 & 44.80 \\ 
SDSS J092014.11$+$453157.3 & 0.402 & 42.62 & 3103 & 20.1 & 9.4 & 45.10 \\ 
SDSS J103951.49$+$643004.2 & 0.402 & 42.99 & 3103 & 27.3 & 21.2 & 45.34 \\ 
SDSS J110621.96$+$035747.1 & 0.242 & 42.71 & 3124 & 5.7 & 0.2 & 44.35 \\ 
SDSS J124337.34$-$023200.2 & 0.281 & 42.60 & 3115 & 7.7 & 2.2 & 44.02 \\ 
SDSS J130128.76$-$005804.3 & 0.246 & 42.83 & 3114 & 6.8 & 1.4 & 44.20 \\ 
SDSS J132323.33$-$015941.9 & 0.350 & 42.77 & 3222 & 11.8 & 6.0 & 44.66 \\ 
SDSS J141315.31$-$014221.0 & 0.380 & 42.83 & 3152 & 10.5 & 7.6 & 44.85 \\ 
SDSS J235818.87$-$000919.5 & 0.402 & 42.90 & 3103 & 19.4 & 11.0 & 44.46 \\ 
\hline
SDSS J091345.49$+$405628.2 & 0.441 & 43.99 & 3005 & 187.1 & 180.9 & 46.38 \\ 

\enddata
\tablecomments{Summary of the 21 sources discussed. Top 11 sources are new \hst\ observations conducted in 2013-2014 (GO-13307, PI Zakamska); middle 9 sources are archival from \hst\ observations conducted in 2003-2004 (GO-9905, PI Strauss); last source was analyzed by \citep{hine99} and we use their results in this paper. $L$[OIII] is given in units of $\log (L{\rm [OIII], erg\, s^{-1}})$, $F_{\nu}$ is the flux density in the $U$-band \hst\ image before and after `optimal subtraction' in $\mu$Jy. $\lambda_{\rm eff}$ is the rest-frame effective wavelength of our $U$-band observations. $L$[12\micron] is given in units of $\log$($\nu L_{\nu}$[12\micron], erg s$^{-1}$).}
\end{deluxetable}

\begin{deluxetable}{l|l|l|l|l|l|l|l|l}
\tablecaption{Model Fit Parameters\label{tab:params}}
\tablehead{ID   & $n_{\rm H,0}$ (cm$^{-3}$) & $n^*_{\rm H,0}$ (cm$^{-3}$) & $\theta (^{\circ})$ & $i(^{\circ})$ & $m$ & max. extent (kpc) & cone-to-countercone }
\startdata
SDSS J014932.53$-$004803.7 & 0.0646 & 0.48 & 26.4 & 54.0 & 0.88 & 7.1 & $>$3.3 \\
SDSS J021047.01$-$100152.9 & 0.0289 & 1.5 & 41.5 & 61.8 & 0.833 & 9.4 & $>$2.3 \\
SDSS J031909.61$-$001916.7 & 0.0417 & 0.47 & 43.1 & 74.3 & 0.772 & 9.8 & $>$1.5 \\
SDSS J031950.54$-$005850.6 & 0.0165 & 1.0 & 22.4 & 43.2 & 0.636 & 7.1 & 5.3 \\
SDSS J032144.11$+$001638.2 & 0.025 & 0.37 & 31.5 & 76.1 & 0.804 & 13.1 & 1.4 \\
SDSS J075944.64$+$133945.8 & 0.00783 & 0.13 & 36.1 & 82.5 & 1.02 & 11.7 & 1.2 \\
SDSS J084130.78$+$204220.5 & 0.127 & 5.4 & 28.2 & 60.6 & 1.61 & 5.7 & 9.6 \\
SDSS J084234.94$+$362503.1 & 0.104 & 39.0 & 41.1 & 70.2 & 1.55 & 10.3 & 1.6 \\
SDSS J085829.59$+$441734.7 & 0.0151 & 0.04 & 34.7 & 70.9 & 0.913 & 8.1 & 1.6 \\
SDSS J103927.19$+$451215.4 & 0.093 & 1.9 & 18.5 & 68.6 & 0.864 & 7.5 & 3.8 \\
SDSS J104014.43$+$474554.8 & 0.123 & 1.4 & 32.1 & 74.8 & 0.898 & 7.4 & 1.5 \\
\hline
SDSS J012341.47$+$004435.9 & 1.1 & 11.7 & 12.5 & 57.6 & 1.07 & 7.2 & $>$2.8 \\
SDSS J092014.11$+$453157.3 & 0.0865 & 0.56 & 20.5 & 63.0 & 0.743 & 6.2 & 2.2 \\
SDSS J103951.49$+$643004.2 & 0.356 & 2.9 & 16.6 & 39.6 & 0.876 & 7.2 & 6.6 \\
SDSS J110621.96$+$035747.1 & 0.00419 & 0.12 & 21.0 & 61.2 & 1.07 & 2.5 & 2.4 \\
SDSS J124337.34$-$023200.2 & 0.00109 & 0.044 & 24.6 & 58.0 & 1.09 & 6.3 & $>$5.7 \\
SDSS J130128.76$-$005804.3 & 0.000536 & 0.051 & 18.4 & 57.6 & 0.987 & 1.0 & 2.7 \\
SDSS J132323.33$-$015941.9 & 0.0244 & 0.52 & 29.6 & 58.3 & 0.782 & 2.9 & 3.0 \\
SDSS J141315.31$-$014221.0 & 0.0119 & 0.12 & 30.5 & 54.0 & 1.02 & 5.1 & 3.3 \\
SDSS J235818.87$-$000919.5 & 0.302 & 6.5 & 14.8 & 47.0 & 1.3 & 6.1 & $>$8.0 \\
\hline
SDSS J091345.49$+$405628.2 & $-$ & $-$ & 23.0 & 37.0 & $-$ & $-$ & $-$ \\ 

\enddata
\tablecomments{Parameters of the density profiles and geometric parameters of the scattering cones derived from our dust scattering models. $\theta$ are half opening angles of scattering cones, $i$ are inclination angles measured relative to the line of sight, $m$ are half-slopes of the density power law profiles, and $n_{\rm H,0}$ is the density at 1 kpc normalized assuming [OIII]-derived bolometric luminosities, whereas $n^*_{\rm H,0}$ values are obtained from 12\micron-derived bolometric luminosities. Maximal projected extents are estimated by looking for $U$-band emission inconsistent with stellar distribution down to limiting surface-brightness of $\lambda' I_{\lambda'}\simeq 3\times 10^{-15}$ erg sec$^{-1}$ cm$^{-2}$ arcsec$^{-2}$. Estimated brightness ratios between forward-scattering and backward-scattering cones are in the last column (lower limits if counter-cone is not detected).}
\end{deluxetable}

\end{document}